\documentclass[12pt]{article}

\usepackage{setspace}
\usepackage{float}
\usepackage[T1]{fontenc} 				
\usepackage{microtype} 				

\usepackage{amssymb,amsmath,amscd}
\usepackage[english]{babel}
\usepackage{titlesec}
\usepackage{hyperref}
\usepackage{graphicx,xcolor}
\usepackage{enumitem}
\usepackage{multirow}
\usepackage[absolute]{textpos}
\usepackage[margin=1.5em,labelfont=bf]{caption}
\usepackage{cite}
\usepackage[bbgreekl]{mathbbol}
\usepackage{bbold}
\DeclareSymbolFontAlphabet{\mathbbl}{bbold}

\hypersetup{
	colorlinks=true,
	linkcolor=dark-blue,
	citecolor=dark-red,
	urlcolor=dark-green,
	linktoc=all
}

\graphicspath{{./figures/}}

\usepackage{geometry} 					
\numberwithin{equation}{section} 

\titleformat{\section}[block]{\Large\bfseries\centering}{\thesection}{1em}{} 
\titleformat{\subsection}[block]{\bfseries}{\thesubsection}{1em}{} 
\titlespacing*{\section}{0pt}{1em}{1em}
\titlespacing*{\subsection}{0pt}{0.75em}{0.75em}

\addto\captionsenglish{
  \renewcommand{\contentsname}%
    {Navigation Guide}%
}

\definecolor{dark-gray}{gray}{0.20}
\definecolor{gray}{gray}{0.30}
\definecolor{light-gray}{gray}{0.80}
\definecolor{dark-red}{rgb}{0.7,0,0}
\definecolor{dark-green}{rgb}{0.1,0.4,0}
\definecolor{dark-blue}{rgb}{0.3,0.3,0.7}
\definecolor{light-blue}{rgb}{0.8,0.8,1}
\definecolor{cardinal}{rgb}{0.6,0,0}
\definecolor{darkgreen}{rgb}{0,0.5,0}
\definecolor{golden}{rgb}{0.92, 0.7, 0}
\definecolor{midnight}{rgb}{0, 0, 0.5}
\definecolor{darkblue}{rgb}{0.2, 0, 0.8}
\definecolor{forestgreen}{rgb}{0.13, 0.55, 0.13}

\def\mop#1{\mathop{\rm #1}\nolimits}

\def\Re{\mop{Re}}
\def\Im{\mop{Im}}
\def\diag{\mop{diag}}

\def\AdS{\mop{AdS}}

\newcommand{\ds}{{\rm d}s}
\newcommand{\dd}{{\rm d}}
\newcommand{\DD}{{\cal D}}
\newcommand{\DDt}{{\widetilde{\cal D}}}
\newcommand{\e}{\mathrm{e}}

\newcommand\Tr{\mathrm{Tr}\,}

\renewcommand\Re{{\rm Re}} 
\renewcommand\Im{{\rm Im}}
\newcommand{\dvol}{{\rm vol}}

\renewcommand{\gcd}{\mathrm{gcd}}
\newcommand{\disc}{\mathbb{D}}
\newcommand{\spindle}{\mathbbl\Sigma}
\newcommand{\p}{\mathfrak{p}}

\newcommand\bN{\mathbf{N}}
\newcommand\bH{\mathbf{H}}

\newcommand\bR{\mathbf{R}}
\newcommand\bZ{\mathbf{Z}}

\newcommand\cA{\mathcal{A}}
\newcommand\cB{\mathcal{B}}
\newcommand\cC{\mathcal{C}}
\newcommand\cD{\mathcal{D}}

\newcommand\cF{\mathcal{F}}
\newcommand\cG{\mathcal{G}}

\newcommand\cL{\mathcal{L}}
\newcommand\cM{\mathcal{M}}
\newcommand\cN{\mathcal{N}}
\newcommand\cO{\mathcal{O}}

\newcommand\cS{\mathcal{S}}

\newcommand\cX{\mathcal{X}}

\newcommand\ff{\mathfrak{f}}
\newcommand\fg{\mathfrak{g}}
\newcommand\fh{\mathfrak{h}}

\newcommand{\f}[2]{\frac{#1}{#2}}

\newcommand{\nn}{\nonumber}

\newcommand {\be} {\begin {equation}}
\newcommand {\ee} {\end {equation}}
\newcommand {\bes} {\begin {equation*}}
\newcommand {\ees} {\end {equation*}}
\newcommand{\lp}{\ell_{\mathrm{p}}}

\newcommand\SL{\mathrm{SL}}

\newcommand\SO{\mathrm{SO}}

\newcommand\UU{\mathrm{U}}
\newcommand\SU{\mathrm{SU}}

\def\overleftrightarrow#1{\vbox{\ialign{##\crcr
			$\leftrightarrow$\crcr\noalign{\kern-0pt\nointerlineskip}
			$\hfil\displaystyle{#1}\hfil$\crcr}}}

\newcommand{\me}{\mathrm{e}}
\newcommand{\ii}{\mathrm{i}}

\title{\fontsize{20pt}{23pt}\selectfont\textbf{Symmetry Breaking and Consistent Truncations\\ from M5-branes Wrapping a Disc}\vspace{5mm}}

\author{Pieter Bomans$\,^{a}$, Christopher Couzens$\,^{a}$, Yein Lee$\,^{b}$, and Sirui Ning$\,^{c}$\\[8mm]
	\normalsize $^a\,$Mathematical Institute, University of Oxford\\
	\normalsize Andrew Wiles Building, Radcliffe Observatory Quarter\\
    \normalsize Woodstock Road, Oxford, OX2 6GG, U.K.\\[2mm]
	\normalsize $^{b}\,$Department of Physics and Research Institute of Basic Science\\
	\normalsize Kyung Hee University, Seoul 02447, Korea\\[2mm]
    \normalsize $^{c}\,$Rudolf Peierls Centre for Theoretical Physics, University of Oxford\\
    \normalsize Beecroft Building, Clarendon Laboratory\\
	\normalsize Parks Road, Oxford, OX1 3PU, U.K.\\[5mm]
	\texttt{\small\href{mailto:pieter.bomans@maths.ox.ac.uk}{pieter.bomans@maths.ox.ac.uk}, \href{mailto:christopher.couzens@maths.ox.ac.uk}{christopher.couzens@maths.ox.ac.uk},}\\
    \texttt{\small\href{mailto:lyi126@khu.ac.kr}{lyi126@khu.ac.kr},
    \href{mailto:sirui.ning@physics.ox.ac.uk}{sirui.ning@physics.ox.ac.uk}}\\
}

\date{}

\begin{document}

\maketitle

\thispagestyle{empty}

\vspace{\stretch{1}}

\begin{abstract}
\noindent We construct new supersymmetric solutions corresponding to M5-branes wrapped on a topological disc by turning on additional scalars in the background. The presence of such scalar fields breaks one of the $\UU(1)$ isometries of the internal space, explicitly realising the breaking by the St\"uckelberg mechanism observed previously. In addition, we construct a consistent truncation of maximal seven-dimensional gauged supergravity on the disc to five-dimensional Romans' $\SU(2)\times \UU(1)$ gauged supergravity, allowing us to construct a plethora of new supergravity solutions corresponding to more general states in the dual SCFTs as well as solutions corresponding to M5-branes wrapping four-dimensional orbifolds.
\end{abstract}

\vspace{\stretch{3}}
 
\newpage
\thispagestyle{empty}

{
    \hypersetup{linkcolor=black}
    \setcounter{tocdepth}{2}
    \tableofcontents
}

\setcounter{page}{0}

\newpage


\section{Introduction and summary}
\label{sec:Introduction}

Strongly coupled quantum field theories have been a subject of intense interest in theoretical physics because they help us to understand fundamental aspects of nature such as, for example, confinement in gauge theories. However, these theories pose a formidable theoretical challenge since traditional perturbative tools become ineffective. Over the last 25 years, holography has offered an alternative method to explore this elusive regime. This duality establishes a connection between certain supersymmetric quantum field theories and higher-dimensional gravity theories and gives valuable insights into the strongly coupled dynamics of the quantum field theories by investigating their dual gravitational descriptions. This remarkable development has opened up various new avenues of research, revealing profound connections between quantum gravity and supersymmetric quantum field theories. As a result, it has become an indispensable tool in understanding some of the most enigmatic aspects of theoretical physics.

A particularly significant set of examples of such strongly coupled field theories are the four-dimensional $\cN=2$ (generalised) Argyres-Douglas theories \cite{Argyres:1995jj,Xie:2012hs}. A large subset of these theories appear at singular points on the moduli space of $\cN=2$ gauge theories, far away from weak coupling, where non-local dyons become simultaneously massless \cite{Argyres:1995xn,Eguchi:1996vu}. For this reason, these theories are intrinsically strongly coupled and evade a conventional Lagrangian approach. Although notoriously hard to access, their existence has been firmly established using both field theoretical and string theoretical arguments. 

A particularly notable construction of these theories involves the compactification of the six-dimensional $\cN=(2,0)$ theory on a sphere with two punctures --- one regular and one irregular \cite{Gaiotto:2009hg,Bonelli:2011aa,Xie:2012hs,Wang:2015mra}. Specifically, within the framework of $(2,0)$ theories of type $A_{N-1}$ this set-up corresponds to M5-branes wrapping this punctured sphere. This M-theory construction introduces a new avenue for investigating these theories via holography. Indeed, the works \cite{Bah:2021mzw,Bah:2021hei} (see also \cite{ Couzens:2022yjl,Bah:2022yjf} for the generalisation to generic regular punctures) argued that these theories can be effectively studied through dual supergravity solutions of the form $\AdS_5 \times \disc$, where $\disc$ topologically is a two-dimensional disc. 

These supergravity backgrounds in question, are similar in nature to those corresponding to M5-branes wrapped on a (possibly punctured) higher genus Riemann surface, \cite{Maldacena:2000mw,Bah:2011vv,Bah:2012dg,Bobev:2019ore} which preserve supersymmetry via a topological twist. One crucial difference arises from the type of metrics that exist on the wrapped two-dimensional surface $\disc$, which necessitates a different way of preserving supersymmetry. The metric on the higher genus Riemann surfaces considered in \cite{Maldacena:2000mw,Bah:2011vv,Bah:2012dg,Bobev:2019ore} have constant curvature, whereas the metric on the disc and its closely related cousin, the spindle, do not admit constant curvature metrics. Consequently, supersymmetry cannot be preserved via a topological twist, but rather using an altogether different mechanism. 

For spindles \cite{Ferrero:2020laf,Ferrero:2020twa,Hosseini:2021fge,Boido:2021szx,Faedo:2021kur,Ferrero:2021wvk,Cassani:2021dwa,Ferrero:2021ovq,Couzens:2021rlk,Faedo:2021nub,Ferrero:2021etw,Couzens:2021cpk,Giri:2021xta,Couzens:2022agr,Cheung:2022wpg,Suh:2022olh,Arav:2022lzo,Couzens:2022yiv,Couzens:2022aki,Boido:2022mbe,Suh:2022pkg,Suh:2023xse,Amariti:2023mpg,Kim:2023ncn,Hristov:2023rel} supersymmetry can be preserved in two ways, the twist or the anti-twist \cite{Ferrero:2021etw}. The twist is topologically a topological twist, with the total R-symmetry flux threading through the spindle precisely cancelling the integrated curvature (Euler characteristic). The anti-twist on the other hand, is even more novel, the total R-symmetry flux through the spindle is not equal to the Euler characteristic, yet supersymmetry is still preserved. Both twists have one crucial ingredient in common, namely that the R-symmetry vector mixes with the isometry of the spindle. In a similar vein, supersymmetry is preserved on the disc by mixing the R-symmetry with the isometry of the disc. The situation on the disc is similar to the anti-twist in that the total R-symmetry flux threading through the disc \cite{Couzens:2021tnv,Suh:2021ifj,Suh:2021aik,Suh:2021hef,Couzens:2021rlk,Karndumri:2022wpu,Couzens:toappear} is not equal to the Euler characteristic. We emphasise, however, that the mechanism differs from the anti-twist and should be seen as a separate mechanism to preserve supersymmetry. One can interpret this mechanism as the twist required to preserve supersymmetry in the presence of an irregular puncture which necessitates the mixing of the sphere isometry with the R-symmetry \cite{Xie:2012hs}. 

Following the success of the spindle and disc solutions, a natural question is whether one can generalise such wrapped brane solutions to higher-dimensional orbifolds. Generalisations of this type have indeed been studied in \cite{Cheung:2022ilc,Couzens:2022lvg,Faedo:2022rqx} and through our work we will be able to further extend these results. 

\paragraph{Summary of results}~

In this paper, we study particular aspects of the holographic duals of (generalised) Argyres-Douglas theories. In the first part of the paper, we comment on a puzzle raised in \cite{Bah:2021hei} regarding the global symmetries of the holographic duals. In the second part of the paper, we construct a consistent truncation of seven-dimensional maximal gauged supergravity on the disc down to five dimensional supergravity allowing us to probe more general states and observables of the dual SCFTs as well as construct new solutions corresponding to M5-branes wrapped on higher-dimensional orbifolds. 

Argyres-Douglas SCFTs generically possess a single $\UU(1)$ global symmetry corresponding to the superconformal R-symmetry.\footnote{Strictly speaking, this is only true when $\gcd(N,k)=1$, where $N$ and $k$ are parameters defining the Argyres-Douglas theory (For more details see \cite{Xie:2012hs}), since for non co-prime $N$ and $k$ the field theory possesses additional global symmetries, however, these additional symmetries are realised holographically in a different way, distinguishing them from the `unwanted' $\UU(1)$ symmetry discussed henceforth.} However, in the holographic duals constructed in \cite{Bah:2021mzw,Bah:2021hei,Couzens:2022yjl,Bah:2022yjf} the supergravity background always contains two $\UU(1)$ isometries. The second isometry can be seen as a remnant of the smearing of the M5-branes over a circle in the internal space. This smearing is in line with the expectation from the Seiberg--Witten geometry which indicates that near the irregular puncture the M5-branes should be intrinsically separated \cite{Wang:2015mra}. This smeared set-up is however merely a remnant of the supergravity description and subleading contributions in $1/N$ are expected to lift the smearing resulting in a localised distribution of (stacks of) M5-branes along the smeared directions. Localising the branes along the smearing circle breaks the additional $\UU(1)$ symmetry, resulting in the correct number of global symmetries. In \cite{Bah:2021hei} the absence of this second $\UU(1)$ was argued for by considering anomaly inflow and the equivariant completion of the four-form flux. They showed that global properties of the solution require the presence of an axion, which through the St\"uckelberg mechanism, gives a mass to one of the gauge fields, breaking the unwanted $\UU(1)$. 

Given that this additional $\UU(1)$ is not an essential ingredient and, stronger even, is supposed to be broken in a bona fide holographic dual it should be possible to explicitly break it and thus give a direct realisation of the proposed St\"uckelberg mechanism. In the first sections of this work we set out to accomplish exactly this. The solutions we will discuss can be obtained by adding a scalar $Y_1$, analogous to the axion in the St\"uckelberg mechanism, parameterising an $\SL(2)/\SO(2)$ coset.\footnote{One can parameterise this coset with one complex scalar. However, one degree of freedom can be absorbed in the gauge field, which ultimately becomes massive. After fixing the gauge, the scalar therefore carries one real degree of freedom.} Similar to the original spindle and disc the local solution we consider can be obtained as an analytic continuation of the BPS bubbling solutions constructed in \cite{Lin:2004nb} and further discussed in \cite{Chong:2004ce}. Our solutions, however, are markedly different when we consider global aspects. Indeed, the straightforward analytic continuation of these bubbling solutions results in a non-compact two-dimensional surface. Such solutions interpolate between a geometry of the form $\AdS_5$ times a conical defect and an asymptotic $\AdS_7$ geometry. Instead of describing a four-dimensional SCFT, they can be interpreted as defects in the six-dimensional $\cN=(2,0)$ theory. This situation is very similar to the set-up considered in \cite{Gutperle:2022pgw,Gutperle:2023yrd} and indeed, using our local solution one can straightforwardly proceed to analyse more general defects within the six-dimensional $\cN=(2,0)$ theory. 

Moving from the local solution to a global completion corresponding holographically to a genuine 4d SCFT, i.e. with a compact two-dimensional internal space $\disc$, proves to be somewhat subtle. Indeed, as we will show it is impossible to find $\cN=1$ backgrounds of this kind with the additional scalars turned on.\footnote{Preserving only $\cN=1$ supersymmetry, one can in principle turn on two scalars $Y_{1,2}$. However, both of them are prohibited when one demands the internal space to be compact.} The local solutions depend crucially on one function $f(w)$, whose roots determine the boundaries of the range of the coordinate $w$ of the two-dimensional internal space.\footnote{There is also another possibility that the space ends at $w=0$ without this necessarily being a root of $f(w)$. We will not comment further on this possibility here.} The (non-)compactness of the internal space $\disc$ is then fully determined by whether this range is (non-)compact. The outcome of our analysis is that compact solutions with an extra scalar only exist when one of the end-points of the line interval is at $w=0$. For the case where this is a root of $f(w)$ the solution simplifies and undergoes supersymmetry enhancement, preserving $\cN=2$ supersymmetry. When the scalar is turned off these backgrounds are exactly the ones described in \cite{Bah:2021mzw,Bah:2021hei} conjectured to be dual to a class of generalised Argyres-Douglas theories. The effect of turning on the scalar $Y_1$ is precisely to break one of the two $\UU(1)$ isometries by providing a mass term for the corresponding gauge field. As such we provide an explicit realisation of the breaking of the unwanted $\UU(1)$ symmetry of the original background. 

In order to support our claims and further analyse our solutions, we uplift them to eleven-dimensional supergravity and embed them into the most general form of $\cN=2$ $\AdS_5$ solutions \cite{Lin:2004nb}. This allows us to compute a range of holographic observables, including central charges, R-symmetry anomalies, and a set of conformal dimensions of operators corresponding to wrapped M2-branes. These observables can be computed on both sides of the duality, and we show that they match exactly in the large $N$ limit. In addition, this allows us to further scrutinise the local internal geometry corresponding to the (ir)regular puncture on the sphere. In particular, we note that near the regular puncture, the scalar field $Y_1$ tends to zero, locally restoring the additional $\UU(1)$ global symmetry. Therefore, close to the puncture, we can use this symmetry to transform the system into an electrostatics problem using a B\"acklund transform \cite{Gaiotto:2009gz} and find local solutions for a more general type of regular puncture. However, unlike the case without the additional scalar field $Y_1$, when moving away from this locus, the second $\UU(1)$ is broken, which implies that we may not perform a global B\"acklund transformation as was used in \cite{Couzens:2022yjl,Bah:2022yjf} in order to construct a global solution with generic regular punctures. 

In the second part of the paper, we shift our focus to a different facet of the Argyres-Douglas theories. In particular, we set out to explore more general solutions of seven-dimensional supergravity by allowing for more general five-dimensional manifolds. Such solutions come in different flavours, some describing more general states in the Argyres-Douglas theories, whereas other describe the compactification of the Argyres-Douglas theory on a generic Riemann surface as well as the compactification of the $\cN=(2,0)$ theory on four-dimensional orbifolds.

The main tool we use to construct these solutions is the development of a consistent truncation of the seven-dimensional maximal $\SO(5)$ gauged supergravity to five-dimensional Romans' $\SU(2)\times \UU(1)$ gauged supergravity \cite{Romans:1985ps}. We construct this truncation by considering the truncation of the LLM geometry to Romans' supergravity considered in \cite{Gauntlett:2007sm} and subsequently specialising their results to our case of interest, namely M5-branes wrapped on a disc both with and without the additional scalar $Y_1$. Similar truncations for M5-branes wrapped on smooth (higher genus) Riemann surfaces and spindles to minimal supergravity have been studied in \cite{Cassani:2020cod,Cheung:2022ilc,Bobev:2022ocx} as well \cite{MatthewCheung:2019ehr} to maximal gauged supergravity in five dimensions.\footnote{Considering the (singular) limit to $\bH^2$, we recover the truncation of \cite{MatthewCheung:2019ehr}.} With this truncation at hand we are now free to take any solution of five-dimensional Romans' supergravity and automatically obtain a corresponding solution of seven-dimensional maximally gauged supergravity. Constructing solutions in the five-dimensional gauged supergravity theory is often much easier than constructing the corresponding wrapped brane solutions directly in seven or eleven dimensions and hence this method gives us access to a wealth of new information.

As a first application one could consider more general asymptotically locally $\AdS_5$ solutions of Romans' supergravity. Through our novel consistent truncation such backgrounds holographically probe more general states in the dual Argyres-Douglas theories, the prime example of this being black hole backgrounds. These backgrounds correspond to the chaotic high energy regime of the dual SCFT dominating the high energy physics. As such, the Bekenstein-Hawking entropy of these supersymmetric black holes reproduces the superconformal index of the dual Argyres-Douglas theories. Indeed, recently there has been a lot of activity studying the thermodynamic and microscopic properties of supersymmetric AdS black holes and their dual SCFT description, see for example \cite{Zaffaroni:2019dhb} and references therein. Following \cite{Bobev:2022ocx}, one can compute the black hole entropy and match it to the large $N$ index of the dual Argyres-Douglas theories. We leave this kind of exploration to a future research endeavour.

A second application of our consistent truncations consists in finding new solutions corresponding to M5-branes wrapped on more general (higher-dimensional) orbifolds. Starting with \cite{Ferrero:2020laf} and subsequent generalisations, it has been appreciated that there are more general ways of preserving supersymmetry on two-dimensional spaces with conical defects. A natural generalisation of this line of thought is to consider higher-dimensional orbifolds. This option has been explored in \cite{Cheung:2022ilc,Faedo:2021nub,Couzens:2022lvg} for M5-branes and D4-branes wrapped on four-dimensional orbifolds. In these works the four-dimensional orbifold has been obtained as a warped products of two spindles. In particular in these cases the lower dimensional theory was always the minimal supergravity theory which only allows for spindle solutions but not discs. 

In this work, we generalise this set-up and instead consider five-dimensional solutions of the form $\AdS_3 \times \disc$ and $\AdS_3\times \spindle$, where $\spindle$ denotes the spindle, and uplift them to seven dimensions to obtain the corresponding $\AdS_3\times\disc\ltimes\disc$ and $\AdS_3\times\spindle\ltimes\disc $ solutions. We discuss in detail the parameter space of such local solutions and show that at generic loci in the parameter space, the disc originating from the seven-dimensional solution is non-trivially fibred over the second disc or spindle. Therefore, the resulting seven-dimensional solutions describe a stack of M5-branes wrapped on a genuine four-dimensional orbifold. On the other hand, for a specific class of $\disc\ltimes\disc$ solutions, the fibration becomes trivial. In this case, the uplifted solution represents a stack of M5-branes wrapped on a factorised four-dimensional space, consisting of two discs $\disc\times\disc$. Finally, our truncation enables the straightforward construction of $\AdS_3\times\Sigma_g\times \disc$ solutions, where $\Sigma_g$ denotes Riemann surface of genus $g$.\footnote{These solutions can be obtained through various (singular) scaling limits of the local solutions presented in Section \ref{subsec:5dsols}.} This, in turn, permits us to employ holography as a means to probe the physics of Argyres-Douglas theories compactified on a generic Riemann surface.

\paragraph{Structure of this paper}~

The remainder of this paper is organised as follows. In Section \ref{sec:SUGRA}, we present the relevant supergravity solutions in seven dimensions. Subsequently, in section \ref{sec:Analysis}, we discuss their uplift to eleven-dimensional supergravity and analyse the novel aspects of our solutions and their impact on the symmetries and anomalies of the dual SCFTs. In section \ref{sec:5dtruncation}, we explore a consistent truncation on the disc and study a selection of novel disc solutions and their implications through the holographic correspondence. We finish with a discussion of future directions in section \ref{sec:Discussion}. Further technical details and a demonstration of the absence of analogous solution for backgrounds corresponding to M5-branes wrapping a spindle are expounded upon in appendices \ref{app:7dSUGRA}-\ref{app:LLM}.

\section{Supergravity solutions}
\label{sec:SUGRA}

We begin by introducing the supergravity solutions of interest and analyse their properties. Generating explicit solutions directly in the context of eleven dimensional supergravity is known to be a difficult task. Therefore, we adopt a two-step strategy: initially constructing these solutions in seven-dimensional gauged supergravity, and subsequently uplifting them to eleven dimensions.

\subsection{Seven-dimensional background}

The seven-dimensional theory of interest is maximal $\SO(5)$ gauged supergravity \cite{Pernici:1984xx} which can be obtained by compactifying eleven-dimensional supergravity on a four-sphere. In order to construct our solutions we will restrict ourselves to a further truncation to $\UU(1)^2$ gauged supergravity. This truncation contains the metric a three-form $C$, two gauge fields $A^{(i)}$, two real neutral scalars $X_i$ and two additional complex scalars $Y_i$, each charged under one of the $\UU(1)$'s.\footnote{The phase of the complex scalars can be removed by a gauge transformation. However, doing so fixes a choice of gauge such that the value of the gauge field at infinity becomes physical.} For more details on these truncations, as well as the BPS equations and equations of motion see Appendix \ref{app:7dSUGRA}.

We are interested in $\AdS_5$ solutions corresponding to M5-branes wrapped on the spindle or disc, as such we choose the following ansatz for the metric,
\begin{equation}\label{eq:7dmetric}
    ds_{7}^{2}=(wH(w))^{\frac{1}{5}}\left(ds_{AdS_{5}}^{2}+\frac{w}{4f(w)}dw^{2}+\frac{f(w)}{H(w)}dz^{2}\right)
\end{equation}
where
\begin{equation}
    \begin{aligned}
        H(w)&=h_{1}(w)h_{2}(w)\\
        f(w)&=\frac{1}{4}H(w)-w^{3}
    \end{aligned}
\end{equation}
The gauge fields and scalars in turn are given by
\begin{equation}\label{eq:7dfields}
    \begin{aligned}
        A^{(i)} = -\f{w^2}{h_i(w)}\,\dd z\,,\qquad X_i(w) = \f{\left(w H(w)\right)^{2/5}}{h_i(w)}\,,\qquad \cosh Y_i(w) = \f1{2w} h^{\prime}_i(w)\,,
    \end{aligned}
\end{equation}
where the prime indicates a derivative with respect to $w$. The three-form necessarily vanishes as a non-zero value is incompatible with the symmetries of our ansatz. One can check that this ansatz indeed solves all equations of motion and BPS equations, provided that the functions $h_i(w)$ solve the following system of non-linear ODEs,\footnote{These solutions can alternatively be obtained as a double analytic continuation of the two-charge black hole solutions of \cite{Chong:2004ce}.}
\begin{equation}\label{eq:ODEsys}
    f(w)\Big( h_i^\prime(w) -w h_i^{\prime\prime}(w) \Big) = w\f{H(w)}{h_i(w)}\left( \f14 h_i^\prime(w)^2-w^2 \right)\,.
\end{equation}
The coordinate $z$ is periodic with period $z \sim z+ 2\pi\Delta z$, while the coordinate $w$ takes value on the interval bounded by two roots of the function $f$. 

Solving the above system of ODEs is prohibitively hard, and we are not able to find a general solution. One particular solution is provided by the functions
\begin{equation}\label{eq:originalSol}
    h_i(w) = q_i+w^2\,.
\end{equation}
Note that in this case the additional scalars vanish and in fact the whole solution reduces to the standard spindle/disc solution as described for example in \cite{Ferrero:2021wvk,Bah:2021mzw,Bah:2021hei}. To find additional non-trivial solutions we proceed by noting that we are looking for functions $h_i(w)$ such that $f(w)$ has at least two real roots and is positive in between. In appendix \ref{app:spindle} we prove that given our ansatz it is not possible to find $\cN=1$ solutions with spindle topology. 

All hope is not lost however, it is possible to find $\cN=2$ solutions. In this case the solution simplifies significantly as the enhanced supersymmetry forces the function $h_2$ to be 
\begin{equation}
    h_2(w) = w^2\,.
\end{equation}
In addition, the gauge fields and scalars reduce to
\begin{equation}\label{eq:N2fields}
    A^{(2)}=0\,, \qquad\qquad X_2(w)=  X_1(w)^{-\f23}\,,\qquad\qquad Y_2(w) =0\,.
\end{equation}
After these simplifications, the ODE \eqref{eq:ODEsys} for $i=2$ is automatically solved, while the remaining non-trivial ODE takes the form,
\begin{equation}\label{eq:N2ODE}
    \left( h_1(W)-4W^{1/2} \right) h_1^{\prime\prime}(W)+\left( h_1^{\prime}(W)^2-1 \right) = 0\,,
\end{equation}
where we changed coordinates, $W=w^2$, and a prime is now understood as a derivative with respect to $W$. Unfortunately, the nature of this equation prevents us from finding a general analytic solution, but we can make progress using numerics, by expanding the solution around the boundaries of the range of $W$; $W=0$ and $W=W_0$ and shooting between the two expansions. 

Around the root, $W=W_0$, we expand the function $h_1$ as follows,
\begin{equation}\label{eq:expansionW0}
    h_1(W) = 4\sqrt{W_0} + (W-W_0) + (W-W_0)^{\alpha} \sum_{k=1}^\infty a_k (W-W_0)^k \,,
\end{equation}
where the first two coefficients are fixed by demanding that $f(W_0)=0$ and $h_1^{\prime}(W_0)=1$ and the exponents $\alpha\in\bN_{>0}$. Substituting this expansion in the ODE \eqref{eq:N2ODE}, we can solve it order by order to obtain a perturbative solution around $W=W_0$. Fixing $\alpha$ to a particular value, the first order determines the position of $W_0$ as\footnote{In order to generalise our ansatz and find a continuous range for the possible values of $W_0$, one could contemplate letting $\alpha$ range over all the positive real numbers. Although this seems to be a sensible option at first, and indeed at first order gives rise to the same condition \eqref{eq:W0cond} for $W_0$, at higher orders this leads to inconsistencies.}
\begin{equation}\label{eq:W0cond}
    W_0 = \f{4\alpha^2}{(2+\alpha)^2}\,.
\end{equation}
This expression indicates that it is not possible to find solutions satisfying the boundary conditions for any value of $W_0$, in particular we find that the maximal possible value of the second root is $W_0=4$. At first this might seem a severe restriction on the possible set of solutions. However, as we will show momentarily, this discrete set includes all of the possibilities allowed by the quantisation of the magnetic flux through the disc. Going to higher order, we find at each order a linear equation determining the coefficients $a_k$ with $k\geq 2$ in terms of the free parameter $a_1$. 

\begin{figure}[!ht]
    \centering
    \includegraphics[width=0.75\textwidth]{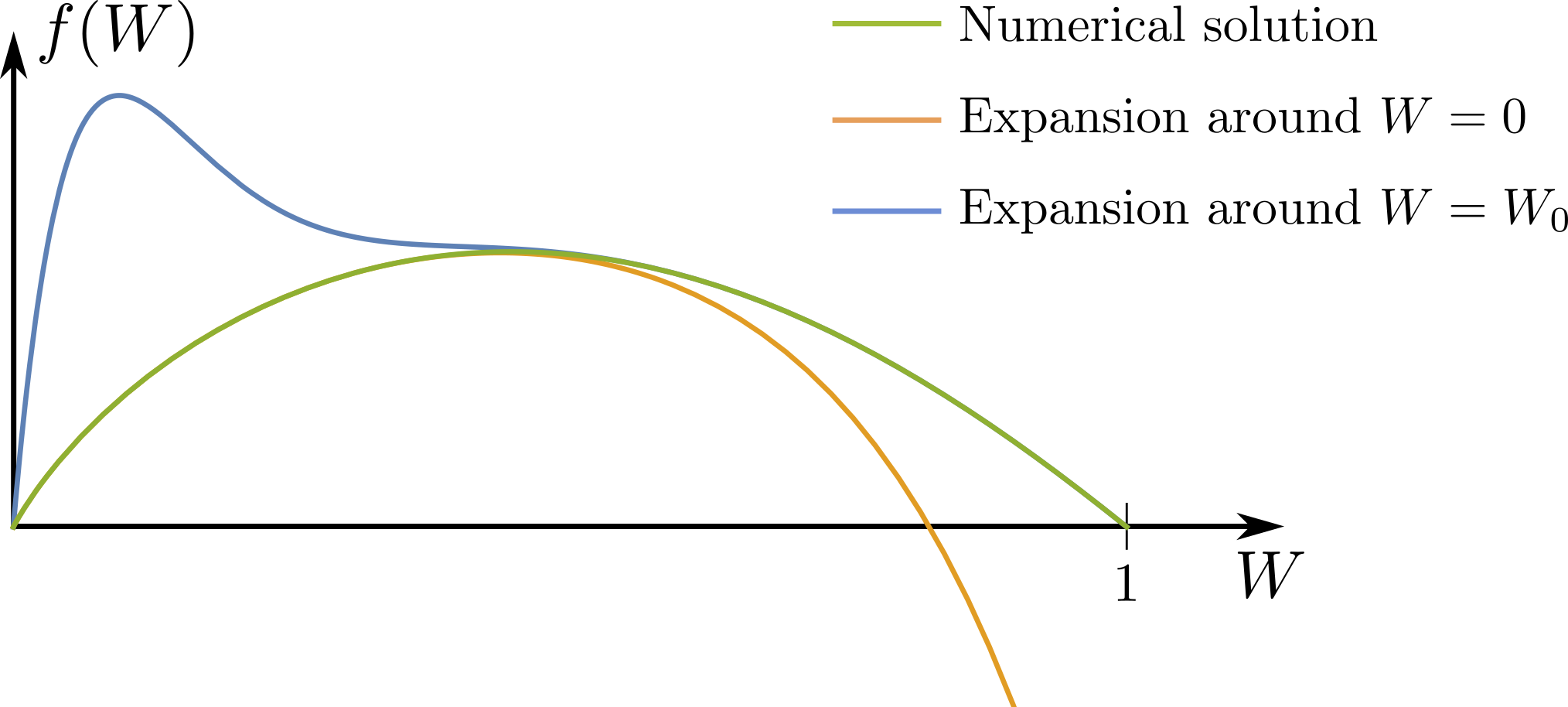}
    \caption{The function $f(W)$ plotted for $\alpha=2$, i.e. $W_0=1$. The parameters in this example are given by $b_0 = 2.5$, $b_2 \simeq 2.4748$ for the expansions around $W=0$ and $a_1 \simeq 0.63$ for the expansion at $W_0=1$.}
    \label{fig:plotf}
\end{figure}

Next, we expand the ODE around $W=0$ at which point an appropriate expansion for the function $h$ is given by, 
\begin{equation}\label{eq:expansion0}
    h_1(W) = \sum_{k=0}^\infty b_k W^{\f k2}\,.
\end{equation}
The first order of the expansion fixes the coefficients $b_1 = b_3 = 0$. Going to higher orders we find that all the coefficients $b_k$ with $k>2$ are determined as a function of the two free parameters $b_0$ and $b_2$. Note that in this expansion we seemingly have one parameter more than in the expansions around $W_0$. However, only one combination of the two coefficients should be fixed in terms of the value of $W_0$ while the other one can be matched with the degree of freedom appearing in the expansion around $W_0$. The value of $W_0$ cannot be directly inferred from the expansion at $W=0$. Instead, we can solve the ODE \eqref{eq:N2ODE} numerically and match the two expansions. From the point of view of the expansion at $W=0$ any value of $W_0$ appears on equal footing. It is only when matching to the expansion at $W_0$ that the condition \eqref{eq:W0cond} enters. We do not have an analytic expression for the map between the parameters of the two expressions. In order to find the map numerically we proceed in two steps. First we find a numerical solution with an allowed value of $W_0$ imposing the boundary conditions at $W=0$. By construction this numerical solution matches the expansion at $W=0$. The second step then consists of numerically minimizing the difference between the expansion at $W=W_0$ and the numerical solution. This procedure then returns the optimal match for the parameter $a_1$. In Figure \ref{fig:plotf} below we demonstrate this matching for $\alpha=2$. The two expansions agree with excellent accuracy with the numerical solution in a large domain. In fact, combining the two expansions completely covers the numerical solution. 

\subsection{Regularity analysis}

The range of the coordinate $w$ is constrained by the requirement that the scalars $X_1$, $Y_1$ be real and the metric be positive definite. Analogous to the disc solutions without the additional scalar we analyse the conditions on the parameters in order to find solutions with orbifold singularities and appropriately quantised magnetic flux \cite{Bah:2021hei,Ferrero:2021etw}. We can study this in general without explicitly solving the ODE as the analysis solely depends on the boundary conditions.

The seven-dimensional metric, \eqref{eq:7dmetric}, takes the form\footnote{For the regularity analysis we go back to the original coordinate $w$. As a reminder, the coordinate transformation to the coordinate $W$ above is given by $W=w^2$.}
\begin{equation}
    ds_{7}^{2}=(wH(w))^{\frac{1}{5}}\left(ds_{AdS_{5}}^{2}+\ds_{\disc}^2\right)\,.
\end{equation}

When approaching $w=w_0$, the internal metric on the disc becomes,
\begin{equation}
    \ds_{\disc}^2 \simeq \f{w_0}{f'(w_0)}\left[ \dd\rho^2 + \f{\rho^2}{n^2} \dd z^2 \right]\,,\qquad\qquad n = -\f{2w_{0}^{2}}{\Delta z\,f'(w_0)}\,,
\end{equation}
where we changed coordinates $\rho = 2(w-w_0)^{1/2}$ and used the two relations $h_1(w_0)=4w_0$ and $h_1^\prime(w_0)=2w_0$ which follow from general properties of the root. Locally around the root $w_0$ the metric therefore takes the form of a $\bR^2/\bZ_n$ orbifold provided that the period of the $z$ coordinate satisfies,
\begin{equation}
    \Delta z=\frac{4}{n(w_0-2)}\,.
\end{equation}
It is useful for later to also introduce the coordinate $z=(\Delta z )\hat{z}$, so that $\hat{z}$ is $2\pi$-periodic.

On the other side of the interval, at $w=0$, we expand the metric using the expansion \eqref{eq:expansion0}, resulting in 
\begin{equation}
    \ds_7^2 \simeq \f{r^{6/5}b_0^{1/5}}{4}\left[ 4\ds_{\AdS_5}^2 + \dd r^2 + \f{16}{b_0} \dd z^2 \right]\,, \qquad\qquad r\rightarrow 0^+\,,
\end{equation}
where we performed the coordinate transformation $w=r^2$. The space $\disc$ has the topology of a disc with a conical defect at $w=w_0$ and a boundary at $w=0$. Indeed, at $w=0$ the $z$ circle does not shrink. The Euler characteristic of the internal space, $\disc$, is given by
\begin{equation}
\begin{aligned}
    \chi(\disc) &=\frac{1}{4\pi}\int_\disc R \,\dvol(\disc) + \frac{1}{2\pi}\int_{\partial\disc}\kappa\, \dvol(\partial\disc)\\
    &=\frac{4\Delta z \,w^{3/2}\big(3 f(w)-w f'(w)\big)}{H(w)^{3/2}}\Bigg|_{w=w_0}\\
    &=\frac{1}{n}\, ,
\end{aligned}
\end{equation}
which is indeed the expected Euler characteristic for a disc with a conical deficit. Note that there is no contribution from the boundary of the disc to the Euler character since the intrinsic curvature $\kappa$ vanishes there.

Next, we consider the quantisation of the magnetic flux through the disc. Due to the presence of the conical defect the flux should be quantised in integer multiples of $\f 1n$. We therefore impose
\begin{equation}
    \frac{\p}{n}=\frac{1}{2\pi}\int_{\mathbb{D}}F^{(1)}=\frac{w_0}{n(2-w_0)} \in \f{1}{n}\bZ \,.
\end{equation}
Hence, the parameter $\p$ should be integer quantised. We can express the (quantised) position of the root as,
\begin{equation}\label{eq:pquant}
    w_0=\frac{2\p}{\p+1}\, ,
\end{equation}
with $\p$ a positive integer. Finally, the holonomy of the gauge field along the boundary remains unchanged from the case without additional scalar and is given by
\begin{equation}
    {\rm hol}_{\partial\disc} \left(A^{(1)}\right) = \oint_{\partial\disc}\f{A^{(1)}}{2\pi} = -\int_{\disc}\f{F^{(1)}}{2\pi} = -\f{\p}{n}
\end{equation}
where we assigned positive orientation to $\dd w\wedge \dd z$.

In the analysis of the ODE above we noticed that the expansion \eqref{eq:expansionW0} around $w=w_0$ is well-behaved only for certain quantised values of the root,
\begin{equation}
    w_0=\frac{2\alpha}{\alpha+2}\, ,\quad \alpha\in \bN_{>0}\, .
\end{equation}
Comparing this with \eqref{eq:pquant} we find that only even values of $\alpha \in 2\bN_{>0}$ are consistent with an appropriately quantised flux. As the value of the magnetic flux increases the power of the leading order term in the expansion of $h_1(w)$ around the solution without the additional scalar similarly increases. In particular, all the allowed values of $\mathfrak{p}$ are included in our expansions and (numerical) solutions. We conclude that for each of the solutions without the additional scalar turned on there is a corresponding family of solutions with the scalar turned on and one additional free parameter given by the value of $Y_1(0)$.

\section{Uplift and analysis}
\label{sec:Analysis}

In order to analyse the $\cN=2$ disc solutions with an extra scalar turned on we proceed to uplift the solution constructed in the previous section to eleven-dimensional supergravity. As we emphasised above, the nature of the ODE \eqref{eq:N2ODE} implies that the solutions only exist for discrete set of parameters. Luckily, this set can be mapped precisely to the set of appropriately quantised solutions where the scalar is trivial and thus for each solution without the additional scalar there exists a deformed solution with the extra scalar. The main goal of this section is to compare and contrast the these two cases by analysing the internal geometry in eleven dimensions and computing a range of holographic observables. 

\subsection{Eleven-dimensional background}

The uplift formulae for general solutions of seven-dimensional $\SO(5)$ gauged supergravity to eleven-dimensional supergravity \cite{Nastase:1999cb,Nastase:1999kf} are given in Appendix \ref{app:Uplift}. Here we restrict ourselves to presenting the final form of the expressions and refer the reader to said appendix for further details. In order to keep the expressions as compact as possible we define the functions,
\begin{equation}
\begin{split}
   Z&=\sin^2\phi\, \me^{Y_1(w)}+\cos^2\phi\, \me^{-Y_1(w)}\,,\\
   \tilde{\Delta}& = w^{4/5} h_1^{3/5}\Delta = \mu^2 h_1(w)+w^2 (1-\mu^2)Z\, ,
    \end{split}
\end{equation}
in terms of which the metric takes the following form,
\begin{align}
    \dd s^2&=\frac{(w\tilde{\Delta})^{1/3}}{4}\bigg[4\dd s^2_{\text{AdS}_5}+\frac{w}{f(w)}\dd w^2+\frac{4 f(w)}{w^2 h_1(w)}\dd z^2+\frac{4\mu^2 w}{\tilde{\Delta}}\dd s^2_{S^2}   \label{eq:11duplift}\\
    &+\frac{4}{\tilde{\Delta}}\bigg(\frac{h_1(w)(1-\mu^2)Z}{w}\Big(D\phi+\frac{2\mu\sin\phi\cos\phi\sinh (Y_1(w))}{(1-\mu^2)Z}\dd\mu\Big)^2+\frac{\tilde{\Delta}}{w(1-\mu^2)Z}\dd\mu^2\bigg)\bigg]\, ,\nonumber
\end{align}
where 
\begin{equation}
D\phi=\dd\phi-A^{(1)}\, .    
\end{equation}
The four-form flux supporting this solution takes the compact form
\begin{equation}
\begin{split}
     G_4&=\dd\bigg[-\frac{\mu^3 h_1(w)}{\tilde{\Delta}}D\phi +\frac{2 w^3 \mu^2 \sin\phi\cos\phi \sinh(Y_1(w))}{\tilde{\Delta}}\dd\mu\bigg]\wedge \dd\text{vol}_{S^2}\, .
\end{split}
\end{equation}
Note that the second term vanishes when the scalar becomes trivial. One can contrast the solution here with the one in \cite{Couzens:2022yjl} where the solution without this additional scalar was presented.\footnote{Different coordinates were used in the two solutions, however the coordinates transformations are straightforward.} The most immediate difference is the presence of the function $Z$ which depends on the coordinate $\phi$. The explicit dependence of the metric on $\phi$ results in the vector field $\partial_{\phi}$ no longer being a Killing vector field of the solution. It is not hard to see that the Lie derivative of the metric along this direction takes the schematic form,
\begin{equation}
\mathcal{L}_{\partial_{\phi}}g_{\mu\nu}\propto \sinh(Y_1(w)) U_{\mu\nu}\,,
\end{equation}
with $U_{\mu\nu}$ a smooth non-vanishing symmetric tensor whose explicit form is not relevant for our discussion. Therefore, on sub-manifolds where $\sinh (Y_1(w)) = 0$, we observe an emergent U$(1)$ symmetry. From the boundary conditions imposed in the previous section we have that at the location of the regular puncture $\sinh (Y_1(w_0))=0$ and thus we find that such enhancement of the symmetry occurs along the surface $w=w_0$. 

\subsection{Regularity analysis}
\label{subsec:regularity}

Next, let us proceed by analysing the regularity of the uplifted metric. One can view the internal space as an $S^2\times S^1$ fibration over a cuboid with the edges defined by the coordinates $(w,\mu, \phi)$. Despite the metric depending on the coordinate $\phi$ it needs to be taken to be $2\pi$-periodic, we will see a consistency check of this fact momentarily. This means that the cuboid should have one face $2\pi$ periodically identified with the mirror face, and is thus $[0,w_0]\times[0,1]\times[0,2\pi)$. Moreover, for all values of $\phi\in[0,2\pi)$ the metric is smooth and therefore we need only consider the degeneration of the internal space along the edges of the rectangle in $(w,\mu)$ coordinates at some fixed value of $\phi$.

\begin{figure}[!ht]
    \centering
    \includegraphics[width=0.6\textwidth]{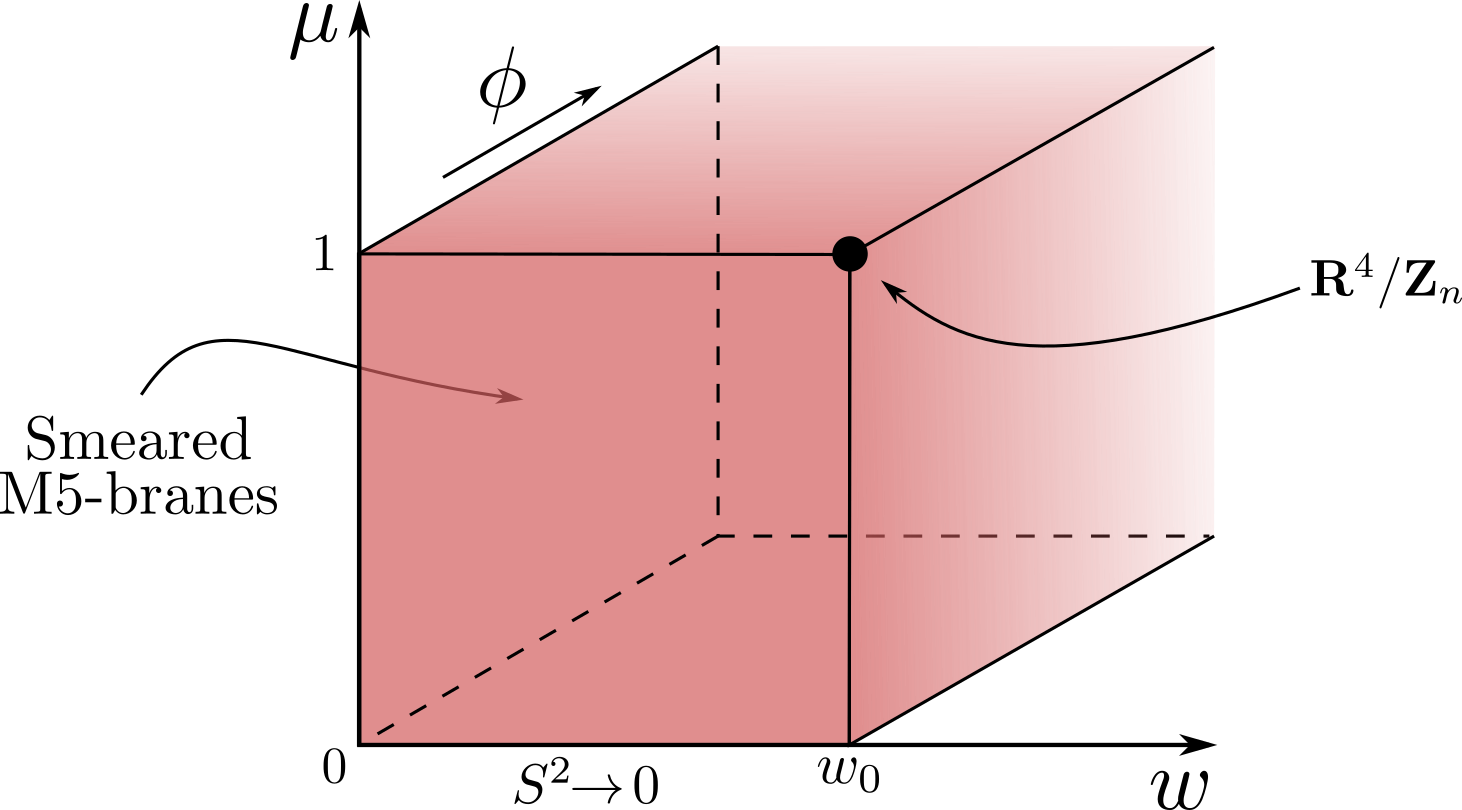}
    \caption{The internal space of the 11d solution is an $S^2\times S^1$ fibration over the cuboid depicted in this figure. At non-zero $\phi$ the metric is regular. At $\phi=\mu=0$, the $S^2$ smoothly shrinks. At $\phi=0$, $\mu=1$ and $w=w_0$ we find a monopole.}
    \label{fig:degenerationCuboid}
\end{figure}

Let us first consider the degeneration at the end-points of the $\mu$ interval. For $\mu=0$ we see that the $S^2$ shrinks smoothly, combining with the $\dd\mu^2$ term to give $\bR^3$ around this point. For the $\mu=1$ degeneration the limit is a bit more subtle. We see that the circle with coordinate $\phi$ shrinks at this point. To see this one should change coordinate as $1-\mu^2=r^2 U(w,\phi)$ with $U(w,\phi)$ a rather complicated function which is nonetheless easily definable as a PDE. Taking the $r\rightarrow 0$ limit one finds that around $\mu=1$ the circle shrinks smoothly giving $\bR^2$. 

We have now studied two of the four edges of the front rectangle in Figure \ref{fig:degenerationCuboid}, it remains to study the metric around $w=0$ and $w=w_0$. First, let us consider the metric around $w=w_0$. Recall that at $w=w_0$ we have $f(w_0)=0$ and $Y_1(w_0)=0$. One finds that the vector
\begin{equation}
    V=n \partial_{\hat{z}}-\p\partial_{\phi}\, ,
\end{equation}
shrinks smoothly along $w=w_0$. At first sight, the conical singularity of the disc at $w=w_0$ seems to have been resolved in the uplifted metric, however this is not fully correct. Indeed, at $w=w_0,\mu=1$ we find a remnant singularity. To see this more clearly we rewrite the metric slightly before changing coordinates adapted to the singular point:
\begin{equation}
    \begin{split}
        \dd s^2=&\frac{(w\tilde{\Delta})^{1/3}}{4}\bigg[4\dd s^2_{\text{AdS}_5}+\frac{w}{f(w)}\dd w^2+\frac{4}{w(1-\mu^2)Z }\dd\mu^2+\frac{4 \mu^2 w}{\tilde{\Delta}}\dd s^2_{S^2}\\
        &+R_z^2(\dd z +L_\phi\dd\phi+L_\mu\dd\mu)^2+R_\phi^2 (\dd\phi+\hat{L}_{\mu}\dd\mu)^2\bigg]\,,
    \end{split}
\end{equation}
where
\begin{equation}
    \begin{split}
        R_z^2&=\frac{4 \big( f(w) \tilde{\Delta}+(1-\mu^2)w^5 Z\big)}{w^2 h_1(w) \tilde{\Delta}}\, ,\quad R_{\phi}^2=\frac{16(1-\mu^2)f(w) Z}{4 w(f(w)\tilde{\Delta}+(1-\mu)^2 w^5 Z}\, ,\\
L_\phi&=\frac{ w^3 (1-\mu^2)h_1(w)Z}{  f(w) \tilde{\Delta}+(1-\mu^2)w^5 Z}\, ,\quad L_{\mu}=\frac{2 \sinh(Y_1(w))w^3 \mu \cos\phi\sin\phi h_1(w)}{f(w) \tilde{\Delta}+(1-\mu^2)w^5 Z}\, ,\quad\\
 \hat{L}_{\mu}&=\frac{2 \sinh(Y_1(w))\mu\cos\phi\sin\phi}{(1-\mu^2)Z}\, .
    \end{split}
\end{equation}
This rewriting shows that the circle with $\phi$-coordinate shrinks for both $\mu=1$ and $w=w_0$ whilst the circle with $z$ coordinate only shrinks at the intersection of the two lines at $w=w_0$ and $\mu=1$. Moreover, observe that $L_\phi$ is piecewise constant along the two edges with\footnote{The expression here differs slightly with that in \cite{Couzens:2022yjl}, this is not an effect of the non-trivial scalar but rather a different choice of gauge for the gauge field which manifests itself in a different definition of the coordinate $\phi$. }
\begin{equation}
    L_{\phi}\Big|_{\mu=1}=0\, ,\quad \frac{L_{\phi}\big|_{w=w_0}}{\Delta z}=-\frac{n}{\p}
\end{equation}
indicating the presence of a monopole at this point. To see this more clearly let us change coordinates as
\begin{equation}
    \mu=1- r^2 \cos^2\tfrac{\zeta}{2}\, ,\qquad w=w_0+\frac{2 f'(w_0)}{w_0^2} r^2 \sin^2\tfrac{\zeta}{2}\, ,\qquad\phi=\hat{\phi}-\frac{z}{2}\, ,\label{eq:w01cof}
\end{equation}
and consider the part of the internal metric excluding the round $S^2$. Taking the limit $r\rightarrow 0$ the four-dimensional metric becomes
\begin{equation}
    \dd s^2_4 \rightarrow \frac{p}{4(p+1)}\bigg[\dd r^2 + \frac{r^2}{4}\Big(\big( n^{-1}d\hat{z}-2 \cos^2 \tfrac{\zeta}{2}\dd\hat{\phi}\big)^2+\dd \zeta^2+\sin^2\zeta \dd\hat{\phi}^2\Big)\bigg]\, ,
\end{equation}
which is precisely the metric on $\bR^4/\bZ_n$. Note that this is analogous with the behaviour of the solution without the scalar turned on, indeed this is to be expected since at this point the scalar necessarily vanishes. 

Finally, we have a single remaining degeneration to consider at $w=0$. This locus located on the left face in Figure \ref{fig:degenerationCuboid} gives rise to a genuine singularity of the solution. However, as in the case without the extra scalar turned on it arises due to the presence of smeared branes \cite{Bah:2021mzw,Bah:2021hei,Couzens:2022yjl}. Carefully taking the $w\rightarrow 0$ limit the metric becomes
\begin{equation}
\begin{split}
\dd s^2&\rightarrow \frac{w^{1/3}\mu^{2/3}h_1(0)^{1/3}}{4}\bigg[4\dd s^2_{\text{AdS}_5}+\dd z^2 +\frac{4}{w \mu^2 h_1(0)}\Big[ \mu^2\big(\dd w^2 +w^2 \dd s^2_{S^2}\big)\\
&+ (1-\mu^2)h_1(0)Z_0\Big(\dd\phi+\frac{2\mu \sinh(Y_1(0))\cos\phi\sin\phi}{(1-\mu^2)Z_0}\dd\mu\Big)^2 +\frac{\mu^2 h_1(0)}{(1-\mu^2)Z_0}\dd\mu^2\Big]\bigg]\, ,
\end{split}
\end{equation}
with
\begin{equation}
    Z_0=\me^{Y_1(0)}\sin^2\phi+\me^{-Y_1(0)}\cos^2\phi\, .
\end{equation}
This is the metric of a stack of M5-branes with world-volume $\AdS_5\times S^1_z$, smeared along two directions. Thus, despite the metric being singular at this point, it degenerates in such a way that is physically sensible. Furthermore, expanding the flux around $w=0$ we find
\begin{equation}
    G_4 \rightarrow \dvol(S^2)\wedge \dd\phi\wedge \dd\mu\, ,
\end{equation}
which is indeed the expected form of the flux for an M5-brane wrapping $\AdS_5\times S^1_z$.

\subsection{Flux quantisation and observables}

Having studied the regularity of the metric we now proceed to appropriately quantise the fluxes and subsequently compute various observables of the dual theory. As a first step, we need to identify all the compact four-cycles in the geometry. The first four-cycle is the $S^4$ from the uplift and may be obtained by fixing $w,z$ to constant values, let us denote this cycle by $\cS$. The second four-cycle arises from considering the circle shrinking at $w=w_0$, $\mu=1$ and the $S^2$ shrinking along $\mu=0$, we denote this one by $\cC$. The final four-cycle consists of the same shrinking circle at  $w=w_0$, $\mu=1$ together with the shrinking $S^2$ along $w=0$. we denote this final four-cycle as $\cD$. 

Considering the first four-cycle $\cS$ we have,
\begin{equation}
 N=   \frac{1}{(2\pi\lp)^3}\int_{\cS}G_4=\frac{1}{\pi \lp^3}\in \mathbb{Z}\,,
\end{equation}
with $N$ the number of M5-branes wrapped on the disc. The two remaining flux quanta are slightly more delicate to compute. Recall that in our analysis of the degeneration at $w=w_0$, $\mu=1$ we were required to perform a coordinate transformation of the angular coordinates, see equation \eqref{eq:w01cof}. Taking this into account we find
\begin{equation}
    \begin{split}
        \frac{1}{(2\pi\lp)^3}\int_{\mathcal{C}}G_4&=\frac{N}{n}\, ,\\
        \frac{1}{(2\pi\lp)^3}\int_{\mathcal{D}}G_4&=\frac{N \mathfrak{p}}{n}\, .
    \end{split}
\end{equation}
In order for the solution to be well defined, we immediately see that we have to require that $\tfrac{N}{n}$ to be integer, while the quantisation of $\mathfrak{p}$ resulting from the 7d solutions ensures that the other flux is appropriately quantised.

\subsubsection{Central Charge}

Having determined the appropriate quantisation conditions, We proceed by computing the central charge of the dual field theory. For an AdS$_5$ solution of the form 
\begin{equation}
    \dd s^2_{11}=\me^{2\lambda}\Big[4 \dd s^2_{\text{AdS}_5}+\dd s^2_{6}\Big]\,,
\end{equation}
the central charge is given by \cite{Gauntlett:2004zh},
\begin{equation}
    a=\frac{2^{5}\pi^3}{(2\pi\lp)^9}\int_{M_6}\me^{9\lambda}\dvol(\cM_6)\, .
\end{equation}
Explicit computation results in, 
\begin{equation}
    a=\frac{N^3 \mathfrak{p}^2}{12 n(\mathfrak{p}+1)}\, .
\end{equation}
Note that result precisely matches the central charge computed for the solution without the scalar turned on. This indeed matches our intuition from the earlier discussion as we conjecture that the sole purpose of this scalar is to break the unwanted isometry, without changing the dual SCFT.

\subsubsection{Conformal dimension of operators}

An additional set of holographic observables that we may compute observables are the dimensions of a set of BPS operators dual to M2-branes wrapped on calibrated two-cycles. The calibrated two-form $\cX$ is given in Appendix \ref{app:LLM}, in terms of which the calibration condition on a 2d sub-manifold $\Sigma_2$ reads,
\begin{equation}
    \cX\big|_{\Sigma_2} = \dvol(\cM_6)(\Sigma_2)\,,
\end{equation}
where the right hand side denotes the restriction of $\cX$ to the world-volume of the probe M2-brane. For such calibrated two-cycles, we can find the dimension of the dual operator to be
\begin{equation}
    \Delta(\Sigma_2) = \f{4\pi}{(2\pi\lp)^3}\int_{\Sigma_2}\e^{3\lambda} \cX\,,    
\end{equation}
where $\lambda$ is a function in the most general $\cN=2$ $\AdS_5$ background as defined in \cite{Lin:2004nb} and is given explicitly in the section below. The calibration form $\cX$ is given in Appendix \ref{app:LLM}. For an M2-brane wrapping the round $S^2$ the calibration condition becomes $y=\me^{-3\lambda}$ which is precisely the location where the R-symmetry vector shrinks. At this locus we find the dimension of the corresponding BPS operator $\cO_1$
\begin{equation}
\Delta(\cO_1)=N\frac{w_0}{2}=N\frac{ \mathfrak{p}}{\mathfrak{p}+1}\, .
\end{equation}
The other choice of calibrated sub-manifold is obtained by considering the $y$ and R-symmetry coordinate located at the north pole of the $S^2$. We find
\begin{equation}
    \Delta(\cO_2)=N\frac{\mathfrak{p}}{n}.
\end{equation}
Similar to the central charges above, we note that there is no modification of these observables in comparison to the solution with the scalar turned off. Similarly, one can show that the entire analysis of symmetries and anomalies of \cite{Bah:2021hei,Couzens:2022yjl,Bah:2022yjf} can be reproduced identically with the only modification being that in this case there is no subtleties in completing the four-form flux into a equivariant form with respect to the isometry corresponding to $\partial_\phi$ since in our background this isometry is explicitly broken.

One can even show that the subleading contributions to the flavour levels, central charges and R-symmetry current algebra levels are also not modified with the inclusion of the additional scalar. We will present results in this direction in \cite{BCtoappear2}.

\section{Consistent truncation on the disc}
\label{sec:5dtruncation}

In the previous sections we have studied a family of disc solutions with a novel scalar turned on. We constructed the eleven-dimensional solutions using the uplift of solutions in a sub-truncation of 7d maximal gauged supergravity. In this section we will show how to perform a consistent truncation of the 7d solution down to 5d Romans' gauged supergravity on the disc. Our consistent truncation has multiple uses, with a key one being that it facilitates the construction of the holographic duals of the Argyres--Douglas theories on an arbitrary Riemann surface. We start this section with a detailed description of the truncation after which we consider a selection of applications.  

\subsection{Embedding 5d supergravity in LLM}
\label{subsec:5dtruncation}

Our method for constructing the consistent truncation from seven to five dimensions by reducing on the disc utilises a number of results in the literature. Rather than explicitly constructing the consistent truncation using (well motivated) trial and error, as in \cite{Cheung:2022ilc,Couzens:2022lvg,Faedo:2022rqx}, we will go through a different route. Our construction uses that there is a known truncation of the Lin--Lunin--Maldacena (LLM) geometries \cite{Lin:2004nb} down to 5d Romans' gauged supergravity. By rewriting our disc solution in the classification of LLM and then reinterpreting the solution as arising from uplifting a 7d solution on the round $S^4$ we may obtain a truncation of the 7d theory on a disc down to 5d Romans' gauged supergravity. Below we present the most important steps of the derivation while referring much of the technical material to appendix \ref{app:LLM}. 

As a first step we briefly review the consistent truncation of the eleven-dimensional LLM geometry to five-dimensional Romans' $\SU(2)\times$U$(1)$ gauged supergravity as worked out in \cite{Gauntlett:2007sm}. The bosonic field content of the five-dimensional gauged supergravity consists of the metric, a real dilatonic scalar field $X$, a $\UU(1)$ gauge field $\cB$, a triplet of $\SU(2)$ gauge fields $\cA^{i}$ and a complex two-form $\cC$ which is charged under the $\UU(1)$ gauge group. The field strengths of the various potentials are defined to be
\begin{equation}\label{eq:Romansfluxes}
\cG_2 = d \mathcal{B}, \qquad
\cF^i_2 = d \mathcal{A}^i - \frac{1}{2\sqrt{2}} \epsilon_{ijk} \mathcal{A}^j \wedge \mathcal{A}^k\,,\qquad
\cF_3 = d\mathcal{C} + \f \ii 2 \mathcal{B} \wedge \mathcal{C}\,.
\end{equation}
In terms or these fields, the eleven-dimensional metric takes the following form
\begin{equation}\label{eq:5dtrunc}
\begin{aligned}
    \ds^2_{11} &= \Big(\frac{\Omega}{X}\Big)^{1/3}\me^{2\lambda}\bigg[4\ds^2_5 +\frac{X \me^{-6 \lambda}}{1-y^2 \me^{-6 \lambda}}\Big(\dd y^2+\me^{D}\big(\dd x_1^2+\dd x_2^2\big)\Big)\\
&+\frac{4 X^2(1-y^2\me^{-6 \lambda})}{\Omega}(\dd\chi+v+\tfrac{1}{2} \mathcal{B})^2 +\frac{y^2\me^{-6 \lambda}}{X \Omega}\DDt\tilde{\mu}^a \DDt\tilde{\mu}^a \bigg]\, ,
\end{aligned}
\end{equation}
where we defined the function
\begin{equation}
\Omega= X y^2\me^{-6 \lambda}+ X^{-2}(1-y^2\me^{-6 \lambda})\, .
\end{equation}
and the gauged one-forms $\DDt\tilde{\mu}$ are given by,
\begin{equation}
    \DDt\tilde{\mu}^a=\dd\tilde{\mu}^a +\frac{1}{\sqrt{2}}\epsilon_{abc}\tilde{\mu}^{b} \mathcal{A}^{c}\,.    
\end{equation}
The expression for the associated four-form flux as well as more details on how to obtain this truncation can be found in Appendix \ref{app:5dLLM}.

\subsection{Embedding 7d supergravity in LLM}

Having defined the truncation from LLM to five dimensional gauged supergravity we proceed to rewrite the seven-dimensional solutions obtained above into the general form of a $\mathcal{N}=2$ AdS$_5$ solution, as classified in \cite{Lin:2004nb}, and reviewed in appendix \ref{app:LLMform}. We can immediately read off both the warp factor and the coordinate $y$,
\begin{equation}
    \me^{6\lambda}= \frac{1}{64} w \tilde{\Delta}\, ,\qquad\qquad y=\frac{w \mu}{4}\, .
\end{equation}
In order to make the R-symmetry vector manifest we define $z=2\chi$ in terms of which we find
\begin{equation}
    v=\frac{1}{\tilde{\Delta}-4 w \mu^2}\Big(4w \mu \sin\phi\cos\phi \sinh(Y_1)\dd\mu+2 w(1-\mu^2) Z \dd\phi\Big)\, .
\end{equation}
At this point we need to distinguish between $Y_1(w)=0$ and $Y_1(w)\neq 0$ to extract out the potential and $x_i$-coordinates.

\paragraph{Non-trivial scalar: $Y_1(w)\neq0$}~

Taking $Y_1(w)$ to be non-zero the $x_i$ coordinates are found to be\footnote{Of course these coordinates are not unique. We are free to perform any $\SO(2)$ transformation on the coordinates as well as perform a constant scaling transformation. The constant rescaling requires that one modifies the Toda potential by an inverse squared power of the scaling parameter.}
\begin{equation}
    x_1=\cos\phi\sqrt{1-\mu^2}\frac{\me^{Y_1(w)/2}}{2\sinh(Y_1(w))}\, ,\quad x_2=\sin\phi\sqrt{1-\mu^2}\frac{\me^{-Y_1(w)/2}}{2\sinh(Y_1(w))}\, .
\end{equation}
The Toda potential in this case is given by
\begin{equation}
    \me^{D}=-\frac{w\sinh^2(Y_1(w))}{8 Y'_1(w)}\,.
\end{equation}
This expression is clearly ill-defined if we take $Y_1(w)=0$.

\paragraph{Trivial scalar: $Y_1(w)=0$}~

When $Y_1(w)=0$ on the other hand we can explicitly solve the ODE \eqref{eq:N2ODE}, finding as solution
\begin{equation}
    h_1(w)=w^2+4(1-a^2)\, .
\end{equation}
In this case, the coordinates $x_i$ are then given by
\begin{equation}
    x_1+\ii x_2=\me^{\ii \phi}\sqrt{1-\mu^2} f_+(w)^{\tfrac{a+1}{2a}}f_-(w)^{\tfrac{a-1}{2a}}\, ,
\end{equation}
where we have defined
\begin{equation}
    f(w)=\frac{w^2}{4}f_+(w)f_-(w)\, ,\qquad f_{\pm}(w)=w-2(1-a)\, ,
\end{equation}
and the associated Toda potential is
\begin{equation}
    \e^{D}=\frac{1}{16} \bigg(\frac{f_-(w)}{f_+(w)}\bigg)^{1/a}\, .
\end{equation}
Note that there is an additional symmetry transformation that may be performed on the coordinates and Toda potential which gives the same metric. Indeed, we can consider the following form of the coordinates and Toda potential
\begin{equation}
\begin{split}
     x_1+\ii x_2&=\me^{\ii k \phi}\Big[\sqrt{1-\mu^2} f_+(w)^{\tfrac{a+1}{2a}}f_-(w)^{\tfrac{a-1}{2a}}\Big]^{k}\,,\\
     \me^{D}&=\bigg(\frac{f_-(w)}{f_+(w)}\bigg)^{1/a}\frac{\Big[\sqrt{1-\mu^2} f_+(w)^{\tfrac{a+1}{2a}}f_-(w)^{\tfrac{a-1}{2a}}\Big]^{2(1-k)}}{16}\, .
     \end{split}
\end{equation}
Given that we accompany the above symmetry transformation with the following shift of the $z=2\chi$ coordinate,
\begin{equation}
    \chi\rightarrow \chi+(1-k)\dd\phi\,,
\end{equation}
the eleven-dimensional metric remains invariant. 

\subsection{Embedding 5d supergravity in 7d supergravity on a disc}

So far we have shown how to embed solutions of both five-dimensional Romans' supergravity as well as seven-dimensional maximal $\SO(5)$ gauged supergravity into the LLM classification. With both these embeddings at hand we are now ready to show how to embed a solution of 5d Romans' theory into the 7d theory on a disc.

Starting from a generic solution of the 5d theory, we can rewrite its embedding into the LLM geometry as an $S^4$ fibration over a 7d space. Doing so we can extract a solution of 7d gauged supergravity from this background following the rules for the embedding of a seven-dimensional solution into the LLM geometry. After some rather tedious manipulations the eleven-dimensional metric as written in \eqref{eq:5dtrunc} takes the form
\small
\begin{align}\label{eq:11dmetric}
    \dd s^2_{11}=&\,\tilde{\Omega}^{1/3}\Bigg[\dd s^2_5+\frac{w X}{f(w)}\dd w^2 +\frac{4 X^4 f(w)}{f(w)+w^3X^3}D\chi^2+\frac{w^2 \mu^2}{16 X^2 \tilde{\Omega}}\DDt\tilde{\mu}_a^2+\frac{4 X}{w(1-\mu^2) Z}\dd\mu^2\\
&+\frac{ \big(f(w)+w^3X^3\big)(1-\mu^2) Z}{4 w^2 X^2\tilde{\Omega}}\Big(\dd\phi+\frac{2\sinh(Y_1)\cos\phi\sin\phi\mu}{(1-\mu^2) Z}\dd\mu-\frac{X^3 w^4}{2(f(w)+w^3 X^3)}D\chi\Big)^2\Bigg]\, ,\nn
\end{align}
\normalsize
where the gauged one-form $D\chi$ is given by
\begin{equation}\label{eq:fibrationcoord}
D\chi\equiv\dd\chi+\frac{1}{2}\mathcal{B}=\frac{1}{2}\big(\dd z+\mathcal{B}\big)\,,
\end{equation}
and we defined the function
\begin{align}
\tilde{\Omega}\equiv&\frac{\Omega}{X\me^{-6 \lambda}}=\frac{4 \mu^2 \big(f(w)+w^3X^3\big)+w^4(1-\mu^2) Z}{64w X^3}\,.\label{eq:Omt}
\end{align}
This metric takes a form very similar to the one we obtained from uplifting a 7d solution to eleven dimensions in \eqref{eq:11duplift}. Indeed, the last step of our derivation consists of interpreting the final four directions of the metric to be those of a squashed $S^4$, so that the 11d metric can be precisely interpreted as the uplift of a 5d solution to 7d maximal $\SO(5)$ gauged supergravity on an $S^4$. At this point it is not hard to see that the fields of 7d maximal gauged supergravity can be extracted as follows. The metric can be written in the canonical form for an uplift from 7d,
\begin{equation}
    \begin{split}
        \dd s^2_{11}=\Delta^{1/3}\bigg[\dd s^2_7+\frac{1}{\Delta}T_{ij}^{-1}\DD\mu^i\DD\mu^j\bigg]\,,
    \end{split}
\end{equation}
with the seven-dimensional metric being
\begin{equation}\label{eq:7d5dmet}
    \dd s^2_7=\frac{ w}{4 X \hat{X}_1^{1/3}}\Big(\dd s^2_5+\frac{w X}{f(w)}\dd w^2 +\frac{4 X^4 f(w)}{f(w)+w^3X^3}D\chi^2\Big)\,.
\end{equation}
The scalar matrix $T_{ij}$ can be read of to be
\begin{equation}
    T_{ij}=\text{diag}\big(\hat{X}_1\me^{-Y_1(w)},\hat{X}_1\me^{Y_1(w)},\hat{X}_2,\hat{X}_2,\hat{X}_2\big)\, ,\label{eq:TijRomans}
\end{equation}
where 
\begin{equation}
 \hat{X}_1=  \bigg[\frac{w^4}{4\big(f(w)+w^3 X^3\big)}\bigg]^{3/5}\, ,\quad \hat{X}_2=\bigg[\frac{4\big(f(w)+w^3 X^3\big)}{w^4}\bigg]^{2/5}\, .
\end{equation}
With a bit more work we can also extract the non-trivial gauge fields from this metric as
\begin{equation}
\begin{split}
   A^{12}&= -\frac{w^4 X^3 }{2 \big(f(w)+w^3 X^3\big)}(\dd \chi+\tfrac{1}{2} \mathcal{B})\, ,\\
   A^{34}&= \frac{1}{\sqrt{2}}\mathcal{A}^{3}\, ,\quad A^{45}= \frac{1}{\sqrt{2}}\mathcal{A}^{1}\,,\quad A^{53}= \frac{1}{\sqrt{2}}\mathcal{A}^{2}\, ,
   \end{split}
\end{equation}
where the second line gives rise to a triplet of $\SU(2)$ gauge fields. 

Finally, to complete our background of 7d gauged supergravity, it remains to extract out the three-form fields $S^i$. These fields do not appear in the metric and hence we need to carefully compare the four-form flux of the five- and seven-dimensional uplift to LLM. This last step is rather technical and the details are referred to Appendix \ref{app:5d7d}. A tedious computation shows that the three-form gauge fields are given by
\begingroup
\allowdisplaybreaks
\begin{align}\label{eq:threeforms}
S^1=&\sqrt{2}\me^{Y_1(w)/2}X^2 \Big[\frac{\sqrt{f(w)}}{w}\star_5 \Im[\me^{-\ii \chi}\mathcal{F}_3]\wedge D\chi -\frac{w}{2\sqrt{f(w)}}\star_5 \Re[\me^{-\ii \chi}\mathcal{F}_3]\wedge \dd w\Big]\nn\\
&-\frac{\sqrt{f(w)}\me^{Y_1(w)/2}}{\sqrt{2}w}\Im[\me^{-\ii\chi}\mathcal{F}_3]\, ,\nn\\
S^{2}=&-\sqrt{2}\me^{-Y_1(w)/2}X^2 \Big[\frac{\sqrt{f(w)}}{w}\star_5 \Re[\me^{-\ii \chi}\mathcal{F}_3]\wedge D\chi +\frac{w}{2\sqrt{f(w)}}\star_5 \Im[\me^{-\ii \chi}\mathcal{F}_3]\wedge \dd w\Big]\nn\\
&+\frac{\sqrt{f(w)}\me^{-Y_1(w)/2}}{\sqrt{2}w}\Re[\me^{-\ii\chi}\mathcal{F}_3]\, ,\nn\\
S^{a+2}=&\frac{w f(w)}{2 \sqrt{2}(f(w)+w^2 X^3)}\mathcal{F}^a\wedge D\chi +\frac{w}{4\sqrt{2} X^2}\star_5 \mathcal{F}^a\, ,
\end{align}
\endgroup
where in the last expression $a\in\{1,2,3\}$. This completes our uplift formulae and as such we have found a complete uplift for any solutions of Romans' supergravity to maximal 7d gauged supergravity.\footnote{Observe that by truncating to minimal gauged supergravity, see appendix \ref{app:5dSUGRA}, we recover (a subset) of the truncation performed in \cite{Cheung:2022ilc}. Note that the gauge choice for the gauge field is uniquely picked for us by our embedding process: the gauge is such that it is the unique R-symmetry vector which is gauged with respect to the $\UU(1)$ graviphoton. This confirms the intuition for why such a gauge choice is important, it is precisely the gauge in which $\partial_z$ is dual to the R-symmetry, see \cite{Cheung:2022ilc,Couzens:2022lvg}.}

\subsection{Applications}
\label{subsec:5dsols}

\begin{figure}[!ht]
    \centering
    \includegraphics[width=\textwidth]{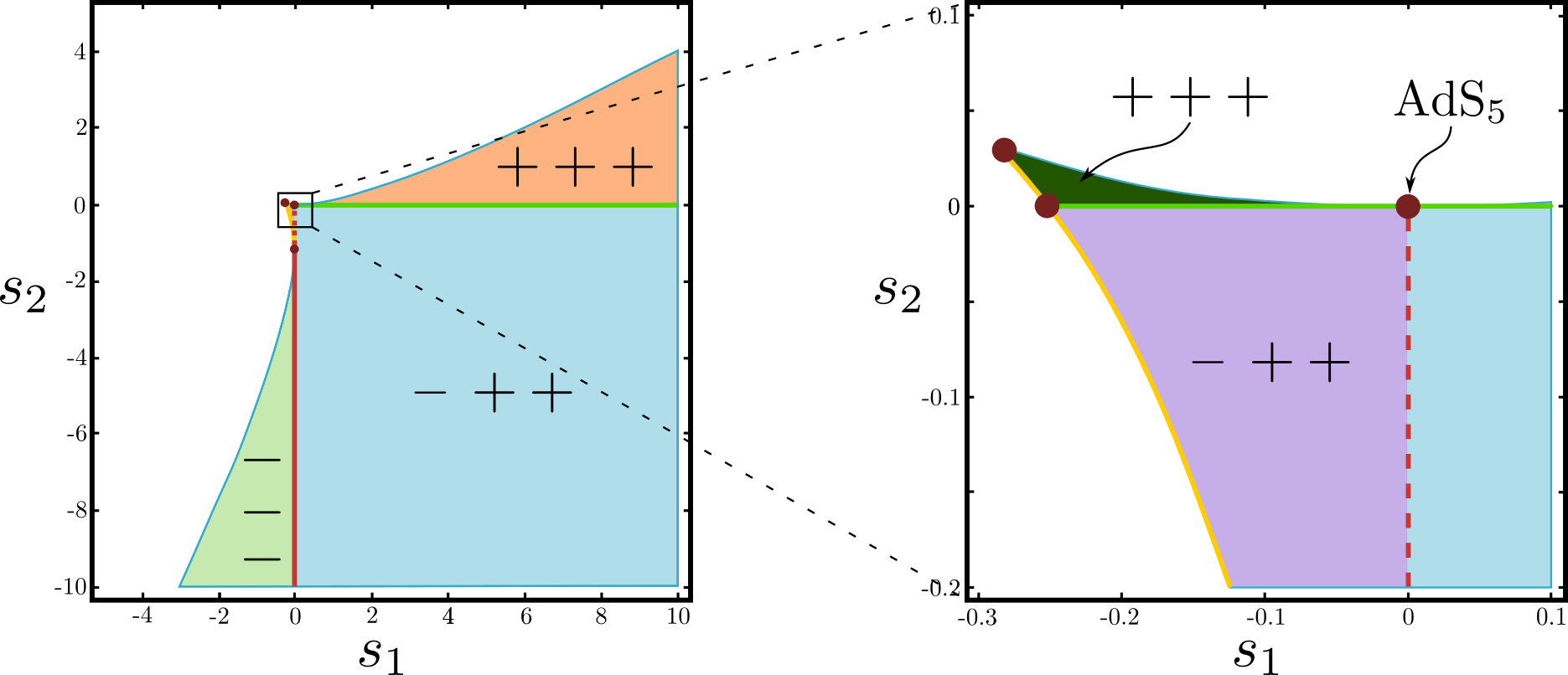}
    \caption{The coloured regions indicate the various allowed regions of the parameter space $(s_1,s_2)$. In each coloured region, the polynomial $\ff$ has three real roots, the signs of which are indicated in the figure. Along the red and green curve the 2d space has the topology of a disc. Along the red curve supersymmetry is enhanced. Along the yellow curve the 2d space contains a cusp-like singurity where the metric locally becomes $\bH^2$. At the intersection of the red and green line the 5d space is given by $\AdS_5$.}
    \label{fig:solmap}
\end{figure}

With the consistent truncation at hand we are now ready to study a variety of new solutions by uplifting solutions from 5d gauged supergravity to seven dimensions. The main examples we consider here are five-dimensional $\AdS_3\times \disc$ and $\AdS_3\times \spindle$ solutions \cite{Hosseini:2021fge,Boido:2021szx}. The local form of the metric is given by
\begin{equation}\label{eq:5dmet}
\dd s^2_5 = \fh(x)^{1/3}\bigg[\dd s_{\AdS_3}^2+\frac{1}{4\ff(x)}\dd x^2+\frac{\ff(x)}{\fh(x)}\dd \psi^2\bigg]\,,
\end{equation}
where we defined the functions,
\begin{equation}
\fh(x)= (x-s_1)^2(x-s_2) \qquad\text{and}\qquad \ff(x)=\fh(x)-x^2\,.
\end{equation}
The remaining non-vanishing fields in this background are given by, 
\begin{equation}\label{eq:5dfields}
\mathcal{B}=\frac{x}{x-s_2}\dd \psi\, ,\qquad \mathcal{A}^3=\frac{x}{x-s_1}\dd\psi\, ,\qquad X=\frac{\fh(x)^{1/3}}{x-s_1}\, .
\end{equation}
Before uplifting this solution to seven dimensions, let us briefly investigate what the global completion of the two-dimensional space represents, depending on the parameters $(s_1,s_2)$. The situation is summarised below and illustrated in Figure \ref{fig:solmap}. 

In the uncoloured region the function $\ff$ only has a single real root and therefore the two-dimensional space is necessarily non-compact. In the coloured regions, $\ff$ has three real roots thus allowing for compact topologies. At a generic point the three roots are distinct and non-zero. The different coloured regions are distinguished by the signs of the roots. As $\ff$ is a cubic monic polynomial, it is positive in between its first two roots. Taking this as the domain of the coordinate $x$ the global solution is given by a compact two-dimensional surface. At such generic points in the allowed region the compact surface has the topology of a spindle and the background preserves $\cN=(2,0)$ supersymmetry. There are however various special loci in the parameter space, at which the global completion has a different topology and/or supersymmetry is enhanced:

\begin{itemize}[itemsep=1.0pt]
    \item $s_1=0$: this locus, indicated by the red line in figure \ref{fig:solmap}, is characterised by trivial $\SU(2)$ gauge fields. In this case, two roots merge to form a double root located at $x=0$. To properly discuss these solutions we need to further divide it into two cases.
    \begin{itemize}[leftmargin=1.5em,itemsep=0pt]
        \item On the sub-locus where $0>s_2>-1$ the bottom two roots merge, and hence compact solutions cease to exist. This is indicated in the figure by the dashed segment of the red line.
        \item For $s_2<-1$, the two largest roots merge and we find a solution with the topology of a disc, analogous to the seven-dimensional backgrounds studied in the first part of this work. 
    \end{itemize} 
    Irrespective of the value of $s_2$, we observe supersymmetry enhancement to (at least) $\cN=(4,2)$ supersymmetry.
    \item $s_2=0$: on this locus, indicated by the green line, the $\UU(1)$ gauge field is trivial. It is distinguished by the smallest root lying at $x=0$. Although at every point where $s_1\neq 0$ this is a single root, the global completion has the topology of a disc. At this locus we observe an enhancement to $\cN=(2,2)$ supersymmetry.
   \item Left boundary segment with $s_2>-1$: along this segment, indicated by the yellow curve, the two largest roots coincide at a value $x_0>0$. Along this edge, $s_2\neq 0$, the two-dimensional space becomes a "black bottle". This geometry has a conical defect at one end-point and a cusp at the other which locally looks like $\bH^2$. Despite the appearance of a seemingly non-compact end-point the space still has finite volume. 
    \item Along the other boundary segments the first two roots coincide hence such solutions cannot give rise to compact surfaces. 
    \item $\left(-\f14,0\right)$: At the intersection of the yellow and green line we have a single root at $x=0$ and a double root at $x_0>0$. The geometry at this point is dubbed a "black goblet" and can be obtained as a limit of the disc where the conical defect becomes a cusp, i.e. locally $\bH^2$.
    \item $(s_1,s_2)=\left(-\f 8{27},\f{1}{27}\right)$: At this intersection of the two boundary segments, $\ff$ has a triple root at $x=\f{4}{27}$. This point does not represent a compact two-dimensional space. 
    \item $(s_1,s_2)=(0,-1)$: At this point we find a triple root at $x=0$. This point does not represent a compact two-dimensional space. 
    \item $(s_1,s_2)=(0,0)$: At this point the five dimensional space reduces to the maximally supersymmetric $\AdS_5$ vacuum of Romans' supergravity with AdS radius $L_{\AdS_5} = \sqrt{\frac{10}{3}}\,$. At this point supersymmetry is enhanced to $\cN=(4,4)$ or, in 4d language, $\cN=2$.
\end{itemize}
Having analysed the various possible global completions of the metric resulting from the local solution \eqref{eq:5dmet} we still have to make sure all fields in the background are well defined. In particular, we have to impose that the scalar has to be positive in the full range of the solution. This restriction, together with the positivity of the metric, imposes the following condition, 
\begin{equation}
    x \geq \max(s_1,s_2)\,,
\end{equation}
between the two smallest roots of the solution. Imposing this constraint rules out the blue, orange and light green regions in Figure \ref{fig:solmap} and leaves us with only the dark green and purple region, including the $\AdS_5$ point and the yellow edge but excluding all other edges. 

We refrain from giving a complete discussion of the flux quantisation in these solutions but rather point out the most important characteristics and leave an in depth discussion of all the different cases for future work. As discussed in \cite{Ferrero:2021etw}, the spindle solutions above can realise the twist or anti-twist in order to preserve supersymmetry.\footnote{In \cite{Ferrero:2021etw}, as well as in \cite{Couzens:2021tnv,Suh:2021ifj}, the compactification of $\cN=4$ SYM on a topological disc is studied. These solutions are therefore constructed in 5d $\UU(1)^3$ gauged supergravity. Romans' supergravity can be obtained by identifying two of the charges in this theory and therefore all their results carry over to our set-up through this specialisation.} Which mechanism is realised depends on the sign of the roots, where two roots of the same sign give rise to a twist, while roots with different signs give rise to the anti-twist. In terms of the colours in Figure \ref{fig:solmap} we see that the purple region gives rise to anti-twist solutions, while the dark green region produces twist solutions. Along the green line we find topological disc solutions preserving enhanced $\cN=(2,2)$ supersymmetry. Similar to the 7d case discussed in section \ref{sec:SUGRA} these solutions realise a distinct yet similar mechanism to the anti-twist to preserve supersymmetry (see for example \cite{Couzens:2021tnv,Suh:2021ifj} for a discussion of the $\cN=(2,2)$ case). The red line is completely excluded except at the point where it intersects with the green line and the geometry reduces to $\AdS_5$. Finally, at the yellow boundary one of the endpoints reduces to a cusp where the metric locally becomes $\bH^2$. In addition, through various (singular) scaling limits, similar to those in appendix C of \cite{Couzens:2022lvg} one can obtain any smooth compact Riemann surface.\footnote{A scaling limit around the point $(s_1,s_2)=\left(-\f8{27},\f1{27}\right)$ gives rise to $T^2$, while a scaling limit at the yellow edge gives rise to compact Riemann surfaces of genus $g>1$. Finally, a scaling limit at the top boundary can give rise to $S^2$ geometries. Observe that in the original solution on which we performed our truncation we may also take a scaling limit to obtain the metric on a constant curvature Riemann surface, in this case only $\mathbb{H}^2$, and necessarily with $Y_1=0$. We therefore obtain solutions of the form $\Sigma_g\times \mathbb{H}^2$, with $g$ arbitrary. These are a subset of the solutions studied in \cite{Benini:2012cz,Benini:2013cda,Faedo:2019cvr}. }

Using our newly constructed consistent truncation, we can uplift the solution \eqref{eq:5dmet} -- \eqref{eq:5dfields} to seven dimensions. One can understand these solutions as the holographic duals of the Argyres--Douglas theories wrapped on the surface $\Sigma$, with metric given in \eqref{eq:5dmet}. 
Inspecting the uplifted metric \eqref{eq:7d5dmet} and \eqref{eq:fibrationcoord} we notice that whenever the $\UU(1)$ gauge field is non-trivial, the disc originating from the 7d solution is non-trivially fibred over the second surface. Such solutions therefore describe novel backgrounds corresponding to a stack of M5-branes wrapping honest four-dimensional orbifolds. The exception are the solutions corresponding to the green line in Figure \ref{fig:solmap}. In this case, the surface has the topology of a disc and supersymmetry is enhanced to $\cN=(2,2)$. For this class of solutions, including the $\AdS_5$ vacuum at $(s_1,s_2)= (0,0)$, the fibration is trivial and the four-dimensional space wrapped by the M5-branes takes a factorised form $\disc_1\times\disc_2$.

We will leave a detailed analysis of the various cases to future work but will finish by making some preliminary comments about the factorised $\disc_1\times\disc_2$ background. To understand in more detail what this solution represents in terms of a brane set-up it proves instructive to analyse it around its various singular points. First, fixing a generic $x \in (0,x_0)$, we see that the solution behaves exactly as the eleven-dimensional solutions described before. Indeed, zooming in around $w\rightarrow0$ and $(w,\mu)\rightarrow (w_0,1)$ we find the same singular behaviour as in section \ref{subsec:regularity}, corresponding to a stack of smeared M5s and a KK-monopole respectively. On the other hand, fixing a generic point $w \in (0,w_0)$, near $x_0$ the solution again behaves as a KK-monopole, but with the roles of the two discs interchanged. Near $x\rightarrow 0$, however, the behaviour is rather different. For generic $w$ the metric at this locus is completely regular. This is in contrast with the result obtained by uplifting the five-dimensional solutions to IIB supergravity where the analogous limit was seen to correspond to D3-branes smeared over three directions \cite{Couzens:2021tnv}. It is only in the simultaneous limit $(x,w,\mu)\rightarrow (0,w_0,1)$ that the metric becomes singular. At this point the metric takes the form of a intersection of a KK-monopole with a stack of (smeared) M5-branes. In addition, one can compute the anomalies of the dual theories using the same tools as in \cite{Benini:2012cz,Benini:2013cda}. While a comprehensive analysis of this singularity is beyond the scope of this project, delving deeper into the intricacies of these singularities would undoubtedly yield intriguing insights.

\section{Discussion}
\label{sec:Discussion}

The objectives of this work are two-fold. We discussed new solutions obtained by adding a scalar $Y_1$ to the existing disc solutions. These solutions explicitly realise the breaking of the unwanted $\UU(1)$ symmetry present in the dual supergravity backgrounds originally presented in \cite{Bah:2021hei,Bah:2021mzw}. The second aim of this paper was to initiate a study of novel solutions corresponding to more general states in general Argyres-Douglas theories as well as solutions corresponding to M5-branes wrapping four-dimensional spaces. These four-dimensional spaces are obtained as the fibration of a disc over a disc or spindle. Obtaining these solutions was made possible through the construction of a consistent truncation of the 7d maximal gauged supergravity on the disc. 

The results presented in this work unfold a plethora of potential avenues for future investigation. To begin with, there is a natural extension of this programme by adapting our results to M2, D3 or D4, wrapping discs or spindles. The local solutions can immediately be established through analytic continuations of the solutions presented in \cite{Chong:2004ce}. It would be interesting to investigate which solutions allow for a global completion as disc or spindle solutions. Of specific interest is the case of D3-branes where the ODEs analogous to \eqref{eq:N2ODE} allow for analytic solutions in a certain limit. In this context, identifying the precise dual SCFTs is an open question. Incorporating an additional set of parameters can provide more refined information aiding this goal.  

Related to the above, in appendix \ref{app:spindle} we discuss the impossibility of adding extra scalars to the solutions corresponding to wrapping M5-branes on a spindle. This indicates that in contrast to the Argyres-Douglas theories, the dual SCFTs have a genuine $\UU(1)^3$ global symmetry which cannot be broken through a St\"uckelberg mechanism. Since in this case the dual SCFT is at present unknown it would be very interesting to further investigate the symmetries in order to make a more informed attempt towards identifying a precise dual SCFT. We are eager to present progress in this direction in the upcoming \cite{BCtoappear}.

In addition, the solutions presented in this work provide novel ways of solving the Toda equation defining an $\cN=2$ $\AdS_5$ background. This example therefore provides a window into obtaining a whole new set of solutions. A key point characterising the solutions in this work is that they break the additional $\UU(1)$ symmetry which was previously present. For this reason, one can no longer globally perform a B\"acklund transform to rewrite the problem as a simpler electrostatics problem. Locally around the regular puncture one is still allowed to do so but finding a consistent way of gluing such local solutions into a global solution remains a outstanding challenge. However, the addition of the novel scalar only mildly affects the structure resulting from the Toda equation. Therefore, it might prove useful to proceed, guided by this example and the strategy of \cite{Bah:2022yjf}, by transforming the Toda equation into a separable form. Doing so successfully could result in a plethora of new solutions without a global axial symmetry. 

Another generalisation of this work is to lift the restriction to compactifications of the six-dimensional theory of type $A_{N-1}$ on a twice punctured sphere. Indeed, the 6d $\cN=(2,0)$ theory comes in different flavours, classifies by a simply laced Lie algebra $\fg\in \{A_{N-1},D_N,E_{6,7,8}\}$. The holographic duals for the Argyres-Douglas theories obtained from six-dimensional type $D_N$ theories as well as the Argyres-Douglas theories obtained by wrapping the $A_{N-1}$ or $D_N$ theories on a sphere with twisted punctures will be presented in \cite{Couzens:toappear}. Starting from their solutions, it becomes a straightforward task to extend our results to encompass this particular case.

Moving away from compact two-dimensional surfaces, one can consider BPS surface defects in the $\cN=(2,0)$ theory --- or any lower dimensional SCFT for that matter. Such defect operators can be represented by exactly the same local solutions. However, to capture the physics of such objects the $w$ coordinate on the disc has to be unbounded such that asymptotically, far away from the defect we find an ambient higher dimensional $\AdS$ region. Such solutions were described in for example \cite{Gutperle:2019dqf,Gutperle:2020rty,Gutperle:2022pgw,Gutperle:2023yrd} in various dimensions. For each of these solutions one can add the scalar described in this work, resulting in more general conformal surface defects. In particular, in $\cN=4$ a well-known class of surface defects is described by Gukov--Witten surface defects \cite{Gukov:2006jk}, which are parameterised by four elements of a maximal torus of the gauge algebra. It is tempting to conjecture that adding additional scalars gives access to a more generic set of these parameters. It would be very interesting to make this proposal more precise.

Finally, in \cite{BenettiGenolini:2023kxp,Martelli:2023oqk}, the authors outlined a method for harnessing the potential of equivariant localisation in supergravity. This was achieved by carefully defining the equivariant action with respect to the $\UU(1)_R$ symmetry present in almost any supersymmetric field theory. When applied to the set-up of LLM geometries, one can utilise their techniques to compute various subleading contributions to the observables computed in this work. Additionally, it offers an elegant and streamlined approach for deriving the anomaly inflow from M-theory \cite{Bah:2018gwc,Bah:2018jrv,Bah:2019jts,Bah:2019rgq}.  Progress made in this direction will be documented in an upcoming publication \cite{BCtoappear2}.

\bigskip
\bigskip
\leftline{\bf Acknowledgements}
\smallskip

\noindent The contributions of PB were made possible through the support of grant No. 494786 from the Simons Foundation and the ERC Consolidator Grant No. 864828, titled "Algebraic Foundations of Supersymmetric Quantum Field Theory" (SCFTAlg). CC is grateful for receiving support by the Mathematics department of the university of Oxford. YL's endeavours were supported by the National Research Foundation of Korea under the grant NRF-2022R1A2B5B02002247. SN received valuable support through a grant from the China Scholarship Council—FaZheng Group at the University of Oxford.

\appendix
\newpage

\section{7d gauged supergravity}
\label{app:7dSUGRA}

In this appendix we clarify our conventions and collect the 7d equations of motion and BPS equations. The relevant supergravity theory is the maximal seven-dimensional $\SO(5)$ gauged supergravity \cite{Pernici:1984xx}. We follow the conventions of \cite{Liu:1999ai}. This theory contains 14 scalars, contained in the $\SL(5,\bR)/\SO(5)$ coset ${V_I}^i$, gauge fields $F^{IJ}$ in the adjoint of $\SO(5)$ and three-form gauge fields $S^I$ in the fundamental of $\SO(5)$. The bosonic Lagrangian of this theory is given by
\begin{equation}\label{eq:lagSO5}
\begin{aligned}
	2\kappa^2 e^{-1} \cL =& R + \f{m^2}{2} (T^2 - 2T_{ij}T^{ij}) - P_\mu{}^{ij}
	P^{\mu\, ij} - \f18 \left(V_I{}^iV_J{}^j F_{\mu\nu}^{IJ}\right)^2 + m^2 \left((V^{-1})_i^I S_{\mu\nu\rho}^I\right)^2\\
    &+ e^{-1}\star\left( \f{m}{48}\,\delta_{IJ}S^I \wedge \dd S^J + \f{1}{16\sqrt{3}} \epsilon_{IJKLM} S^I \wedge F^{JK}\wedge F^{LM} + m^{-1} CS_7(A) \right)\,,\\
\end{aligned}
\end{equation}
where $CS_7(A)$ denotes the seven-dimensional Chern-Simons functional for the $\SO(5)$ principal bundle. For all situations of interest in this paper $CS_7(A)$ vanishes and hence we will ignore this term from here on. The tensor $T_{ij}$ is defined as $T_{ij} = V^{-1}{}_i{}^IV^{-1}{}_j{}^J \delta_{IJ}$ and $T=\Tr T$. Both $I,J=1,\dots 5$ and $i,j=1,\dots 5$ indices are raised and lowered with the Kronecker delta. The scalar kinetic term is defined in terms of $P^{ij}_\mu$, which is given as the symmetric part of $V^{-1}{}_i{}^I \cD_\mu V_I{}^j$ where $\cD_\mu V_I{}^i = \partial_\mu V_I{}^i + A_\mu{}_I{}^J V_J{}^i$.

In this paper we consider a $\UU(1)^2$ truncation of this theory where the only non-vanishing gauge fields are given by $F^{(1)} = F^{12}$ and $F^{(2)} = F^{34}$. The scalar manifold in our truncation is given by 
\begin{equation}
    \SO(2)^2\times \left(\f{\SL(2,\bR)}{\SO(2)}\right)^2
\end{equation}
Where the $\SO(2)$ factors parameterise two scalars $X_i$ neutral under the $\UU(1)^2$ gauge symmetry and the $\f{\SL(2,\bR)}{\SO(2)}$ cosets parameterise two charged scalars $Y_i$. In terms of the coset $V_I^j$ the scalars in our truncation are given by
\begin{equation}
	V_I^i = \diag\left[ X_1^{-1/2}\e^{Y_1/2}\,, X_1^{-1/2}\e^{-Y_1/2} \,,
	X_2^{-1/2}\e^{Y_2/2} \,, X_2^{-1/2}\e^{-Y_2/2} \,, X_1 X_2 \right]\,.
\end{equation}
Although the presence of an external three-form is incompatible with the symmetries of our ansatz in section \ref{sec:SUGRA}, they will be needed when we consider more general solutions obtained from our novel consistent truncation.

Substituting the above truncation in \eqref{eq:lagSO5}, we obtain the following Lagrangian
%
\begin{align}
	2\kappa^2 e^{-1} \cL =& R - V - 5\partial(\lambda_1+\lambda_2)^2 -
	\partial(\lambda_1-\lambda_2)^2 - \f12 \partial Y_1{}^2- \f12
	\partial Y_2{}^2  \nn\\
	&-2 \left(  \sinh^2 Y_1\, A^{(1)2} + \sinh^2 Y_2\,A^{(2)2}  \right) - \f12\e^{-4\lambda_1}F^{(1)}{}^2-
    \f12\e^{-4\lambda_2}F^{(2)}{}^2\\
 &+m^2e^{-4\lambda_1-4\lambda_2}S^2+e^{-1}\star\left(\f{m}{6} 
 S\wedge \dd S+\f{1}{2\sqrt{3}}\, S\wedge F^{(1)}\wedge F^{(2)}\right)\,,\nn
\end{align}
where the potential $V$ is given by
\begin{equation}
\begin{aligned}
	V = \f{m^2}{2}\e^{-8(\lambda_1+\lambda_2)}\Big(&
	2+4\e^{12\lambda_1+8\lambda_2}\cosh2Y_1+4\e^{8\lambda_1+12\lambda_2}\cosh2Y_2\\
	&\,\, - \left(
	1+2\e^{6\lambda_1+4\lambda_2}\cosh Y_1+2\e^{4\lambda_1+6\lambda_2}\cosh Y_2
	\right)^2 \Big)\,.
\end{aligned}
\end{equation}
and the length scale of the $\AdS_7$ vacuum of this theory is given by $L_{\AdS}=\f{2}{m}$.\footnote{In the main text we set $m=1$ but it can be reinstated straightforwardly.} For future convenience, we redefined $X_i = \e^{2\lambda_i}$. 

The equations of motion derived from this Lagrangian are as follows. The scalar equations are given by
\begin{equation}
\begin{aligned}
    \nabla^2(3\lambda_1+2\lambda_2) =& -\f14\e^{-4\lambda_1}F^{(1)}{}^2  + m^2e^{-4\lambda_1-4\lambda_2}S^2+\f14
	\partial_{\lambda_1}V\,,\\
	\nabla^2(2\lambda_1+3\lambda_2) =& -\f14\e^{-4\lambda_2}F^{(2)}{}^2+ m^2e^{-4\lambda_1-4\lambda_2}S^2 + \f14
	\partial_{\lambda_2}V\,,\\
	\nabla^2(Y_1) =& 2\sinh^2 Y_1 \,A^{(1)2} + \partial_{Y_1}V\,,\\
	\nabla^2(Y_2) =& 2\sinh^2 Y_2 \,A^{(2)2} + \partial_{Y_2}V\,.
\end{aligned}
\end{equation}
The gauge field equations of motion are given by
\begin{equation}
\begin{aligned}
	\nabla^\mu(\e^{-4\lambda_1}F^{(1)}_{\mu\nu}) =& 4 \sinh^2 Y_1 \, A^{(1)}_\nu+\frac{1}{2\sqrt{3}}\epsilon_{\mu\nu}{}^{\lambda\sigma\alpha\beta\gamma}\nabla^\mu(F_{\lambda\sigma}^{(2)}S_{\alpha\beta\gamma}),\\
	\nabla^\mu(\e^{-4\lambda_2}F^{(2)}_{\mu\nu}) =& 4\sinh^2Y_2\, A^{(2)}_\nu+\frac{1}{2\sqrt{3}}\epsilon_{\mu\nu}{}^{\lambda\sigma\alpha\beta\gamma}\nabla^\mu(F_{\lambda\sigma}^{(1)}S_{\alpha\beta\gamma}).
\end{aligned}
\end{equation}
The three-form $S$ satisfies the following self-duality equation
\begin{equation}
    \e^{-4\lambda_1-4\lambda_2}S_{\mu\nu\rho} = \f1{6m}\epsilon_{\mu\nu\rho}{}^{\alpha\beta\gamma\delta}\partial_\alpha S_{\beta\gamma\delta} - \f{1}{8\sqrt{3}m^2}\epsilon_{\mu\nu\rho}{}^{\alpha\beta\gamma\delta} F^{(1)}_{\alpha\beta}F^{(2)}_{\gamma\delta}\,.
\end{equation}
Finally, the (trace subtracted) Einstein equation is given by 
\begin{equation}
\begin{aligned}
    R_{\mu\nu} =& \f15 g_{\mu\nu}V +
	5\partial_\mu(\lambda_1+\lambda_2)\partial_\nu(\lambda_1+\lambda_2) +
	\partial_\mu(\lambda_1-\lambda_2)\partial_\nu(\lambda_1-\lambda_2) \\
	&+ \f12 \partial_\mu Y_1\partial_\nu Y_1 + \f12
	\partial_\mu Y_2\partial_\nu Y_2 +2\sinh^2 Y_1\, A_\mu^{(1)}A_\nu^{(1)} + 2\sinh^2 Y_2 \,A_\mu^{(2)}A_\nu^{(2)} \\
	&+ \f12 \e^{-4\lambda_1}\left( F^{(1)}_{\mu\nu}{}^2 - \f1{10}g_{\mu\nu}F^{(1)}{}^2
	\right) + \f12 \e^{-4\lambda_2}\left( F^{(2)}_{\mu\nu}{}^2 -
	\f1{10}g_{\mu\nu}F^{(2)}{}^2 \right)\\
    &-3m^2e^{-4\lambda_1-4\lambda_2}(S^2_{\mu\nu}-\frac{2}{15}S^2)\,.
\end{aligned}
\end{equation}
The supersymmetry variations are given by 
\begin{align}
	\delta\psi_\mu =& \bigg[ \cD_\mu + \f{m}{20}T\gamma_\mu - \f{1}{80}\left(
	\gamma_\mu{}^{\nu\rho} -8\delta_\mu^\nu\gamma^\rho \right) \Gamma^{ij}		
    V_I^iV_J^j F_{\nu\rho}^{IJ}\\
    &\qquad\qquad\qquad+\f{m}{10\sqrt{3}}\left( \gamma_\mu{}^{\nu\lambda\sigma}-\f92 \delta_\mu^\nu \gamma^{\lambda\sigma} \right)\Gamma^i V^{-1}{}_i^I S^I_{\nu\lambda\sigma}\bigg]\epsilon\,,\nn\\
	\delta\lambda_i =& \Bigg[ \f{m}{2}\left(T_{ij}-\f15
	\delta_{ij}T\right)\Gamma^j +\f12 \gamma^\mu P_{\mu ij}\Gamma^j
    + \f{1}{32}\gamma^{\mu\nu}\left( \Gamma^{kl}\Gamma^i -\f15 \Gamma^i\Gamma^{kl}
	\right) V_K^kV_L^l F_{\mu\nu}^{KL}\\
    &\qquad\qquad\qquad+\f{m}{20\sqrt{3}}\gamma^{\mu\nu\lambda}\left( \Gamma^{ij} - 4\delta^{ij} \right) V^{-1}{}_j^J S^J_{\mu\nu\lambda}\Bigg]\epsilon\,.\nn
\end{align}
They simplify considerably when we consider the following linear combinations,
\begin{align}
	\hat\psi_\mu &= \psi_\mu + \f12 \gamma_\mu \Gamma^5\lambda_5\,,& & \\
	\lambda^{(1)} &= (\Gamma^1\lambda_1+\Gamma^2\lambda_2) + \f32
	(\Gamma^3\lambda_3+\Gamma^4\lambda_4)\,,& \lambda^{(3)} &=
	(\Gamma^1\lambda_1-\Gamma^2\lambda_2)\,,\\ \lambda^{(2)} &=
	\f32(\Gamma^1\lambda_1+\Gamma^2\lambda_2) + 
	(\Gamma^3\lambda_3+\Gamma^4\lambda_4)\,,&
	\lambda^{(4)} &= (\Gamma^3\lambda_3-\Gamma^4\lambda_4) \,.
\end{align}
For the truncation under consideration the supersymmetry variations reduce to the set following equations,
\begingroup
\allowdisplaybreaks
\begin{align}
	\delta \hat\psi_\mu =& \Big[ \nabla_\mu + \f{g}{2}\left(
	A^{(1)}_\mu\Gamma^{12}+A^{(2)}_\mu\Gamma^{34} \right) +
	\f{m}{4}\e^{-4(\lambda_1+\lambda_2)}\gamma_\mu \label{eq:psivar}\\
	&+ \f14 \gamma^\nu \left(
    \e^{-2\lambda_1}F^{(1)}_{\mu\nu}\Gamma^{12}+\e^{-2\lambda_2}F^{(2)}_{\mu\nu}\Gamma^{34}
	\right)
	+\f12 \gamma_\mu\gamma^\nu \partial_\nu (\lambda_1+\lambda_2) \nn\\
    &-\frac{m\sqrt{3}}{4}\gamma^{\nu\lambda}e^{-2\lambda_1-2\lambda_2}S_{\mu\nu\lambda}\Gamma^5\Big] \epsilon
	\nn\\
	\delta \lambda^{(1)} =& \Big[ \f{m}{2}\left(
	\e^{2\lambda_1}\cosh Y_1-\e^{-4(\lambda_1+\lambda_2)} \right) - \f12
	\gamma^\mu\partial_\mu (3\lambda_1+2\lambda_2) - \f18
	\gamma^{\mu\nu}\e^{-2\lambda_1}F_{\mu\nu}^{(1)}\Gamma^{12}\label{eq:l1var}\\
    &+\frac{m}{4\sqrt{3}}\gamma^{\mu\nu\lambda}e^{-2\lambda_1-2\lambda_2}S_{\mu\nu\lambda}\Gamma^5\Big] \epsilon \nn\\
	\delta \lambda^{(2)} =& \Big[ \f{m}{2}\left(
	\e^{2\lambda_2}\cosh Y_2-\e^{-4(\lambda_1+\lambda_2)} \right) - \f12
	\gamma^\mu\partial_\mu (2\lambda_1+3\lambda_2) - \f18
	\gamma^{\mu\nu}\e^{-2\lambda_2}F_{\mu\nu}^{(2)}\Gamma^{34}\label{eq:l2var}\\
    &+\frac{m}{4\sqrt{3}}\gamma^{\mu\nu\lambda}e^{-2\lambda_1-2\lambda_2}S_{\mu\nu\lambda}\Gamma^5\Big] \epsilon \nn\\
    \delta \lambda^{(3)} =& \left[m\e^{2\lambda_1}\sinh Y_1 - \f12
	\gamma^\mu\partial_\mu Y_1 + \sinh Y_1 \gamma^\mu A^{(1)}_\mu \Gamma^{12}
	\right]\epsilon\label{eq:l3var}\\
	\delta \lambda^{(4)} =& \left[m\e^{2\lambda_2}\sinh Y_2 - \f12
    \gamma^\mu\partial_\mu Y_2 + \sinh Y_2 \gamma^\mu A^{(2)}_\mu
	\Gamma^{34}\right]\epsilon\,.\label{eq:l4var}
\end{align}
\endgroup
As usual, supersymmetry fixes the gauge coupling in terms of the $\AdS$ length as $g=2m$. Having introduced the general equations of motion and BPS equations, in the following we will restrict ourselves to the $\AdS_5$ solutions introduced in Section \ref{sec:SUGRA}. As already noted before, the three-form necessarily has to vanish for this type of solution. The metric and gauge fields take the form
\begin{equation}
	\ds^2 = f(w)\ds_{\AdS_5}^2 +  g_1(w) \dd w^2 + g_2(w) \dd z^2\,, \qquad\qquad A^{(i)} = A^{(i)}_z(w)\,\dd z\,,
\end{equation}
and all scalars solely depend on the coordinate $w$. 

With this ansatz at hand we can now proceed to further simplify the BPS equations following a similar approach as in \cite{Bah:2021hei}. To do so it is useful to introduce an explicit representation of the 7d gamma matrices,
\begin{equation}
	\gamma^{\alpha} = \rho^{\alpha}\otimes \sigma^3 \,,\qquad \gamma^6 = 1_4
	\otimes \sigma^1\,,\qquad \gamma^7 = 1_4 \otimes \sigma^2\,,
\end{equation}
where $\alpha=1,\dots,5$ runs along the $\AdS_5$ directions. The five dimensional gamma matrices $\rho^\alpha$ are given by 
\begin{equation}
    \begin{aligned}
        \rho^1&=i \sigma^1\otimes 1_2\,,\qquad& \rho^2&=\sigma^2\otimes 1_2\,,\qquad&\rho^3&=\sigma^3\otimes \sigma^1\,,\\
        \rho^4&=\sigma^3\otimes \sigma^2\,,&
        \rho^5&=\sigma^3\otimes \sigma^3\,.
    \end{aligned}
\end{equation}
In line with the $5+2$ split of the gamma matrices we write the supersymmetry parameter as a tensor product,
\begin{equation}
	\epsilon^I = n^I \theta \otimes \eta\,,
\end{equation}
where $\theta$ is a conformal Killing spinor on $\AdS_5$ satisfying,
\begin{equation}
	\nabla_\alpha^{\AdS_5}\theta = \f12 s_1 \rho_\alpha \theta\,, \quad s_1=\pm
	1\,,
\end{equation}
where the sign $s_1$ is arbitrary. For concreteness we fix it to be $s_1=1$ in the remainder of the analysis. The constants $n^I$ are the components of an object in the $\mathbf{4}$ of $\SO(5)$ transforming under the $\SO(5)$ $\Gamma$-matrices. The 2-component spinor $\eta$ depends only on $w$ and $z$ and moreover, since $\partial_z$ is a Killing vector we can assume that the spinor has a definite charge under the $\UU(1)_z$ isometry. In other words we have
\begin{equation}
    \eta(w,z) \equiv \e^{i q z}\eta(w)\,.
\end{equation}
Depending on the solution, we have to impose either one or two projection conditions corresponding to respectively a 4d $\cN=2$ or $\cN=1$ dual SCFT. In this appendix we analyse the more general case with $\cN=1$ supersymmetry. When a solution preserves $\cN=2$ supersymmetry one of the projectors (or a linear combination thereof) will be redundant and does not need to be imposed. For minimally supersymmetric solutions, the projection conditions are as follows,
\begin{equation}
	(\Gamma^{12})^I{}_J n^J = i n^I\,,\qquad (\Gamma^{34})^I{}_J n^J = i n^I
	\,.
\end{equation}
in principle we should allow for arbitrary signs in these equations but these can always be absorbed in the gauge fields. 

Substituting the explicit form of the gamma matrices and factorised spinors in the BPS equations \eqref{eq:psivar}-\eqref{eq:l4var} we find six algebraic equations,
\begingroup
\allowdisplaybreaks
\begin{align}
	0 =& \f{s_1}{f^{1/2}}\eta + \f{1}{g_1^{1/2}}  \left( \f{f^\prime}{2f} +
	\lambda_1^\prime+ \lambda_2^\prime \right)(i\sigma^2\eta)+\f m2 
	\e^{-4(\lambda_1+\lambda_2)}(\sigma^3\eta)\,, \\
	0 =& - \f1{2(g_1g_2)^{1/2}} \left(
	\e^{-2\lambda_1}A^{\prime(1)}_z + \e^{-2\lambda_2}A^{\prime(2)}_z\right)\eta + \f g{2  g_2^{1/2}}\left(A^{(1)}_z+A^{(1)}_z+\f{2q}{g}\right)(\sigma^1\eta) \nn\\
	& + \f{1}{g_1^{1/2}}  \left( \f{g_2^\prime}{2g_2} + \lambda_1^\prime +
	\lambda_2^\prime \right)(i\sigma^2\eta)
	+\f{m}{2}\e^{-4(\lambda_1+\lambda_2)}(\sigma^3\eta)\,,\\
	0 =& \f1{2(g_1g_2)^{1/2}}\e^{-2\lambda_1}A^{\prime(1)}_z \eta -
	\f{1}{g_1^{1/2}}(3\lambda_1^\prime+2\lambda_2^\prime)(i\sigma^2\eta) +  m\left(
	\e^{2\lambda_1}\cosh Y_1-\e^{-4(\lambda_1+\lambda_2)} \right)(\sigma^3\eta)
	\,, \\
	0 =&  \f1{2(g_1g_2)^{1/2}}\e^{-2\lambda_2}A^{\prime(2)}_z\eta -
	\f1{g_1^{1/2}}(2\lambda_1^\prime+3\lambda_2^\prime)(i\sigma^2\eta) +  m\left(
	\e^{2\lambda_2}\cosh Y_2-\e^{-4(\lambda_1+\lambda_2)} \right)(\sigma^3\eta)
	\,, \\
	0 =& \f1{g_2^{1/2}}\sinh Y_1 A^{(1)}_z \left( \sigma_1\eta \right) + \f1{2g_1^{1/2}} Y_1^\prime(i\sigma^2\eta) -
	m\e^{2\lambda_1}\sinh Y_1(\sigma^3\eta)\, ,\\	
    0 =&\f1{g_2^{1/2}}\sinh Y_2 A^{(2)}_z \left( \sigma_1\eta \right) + \f1{2 g_1^{1/2}} Y_2^\prime(i\sigma^2\eta) -
	m\e^{2\lambda_2}\sinh Y_2(\sigma^3\eta)\,.
\end{align}
\endgroup
In addition, the $w$ component of \eqref{eq:psivar} gives rise to one ODE which determines the Killing spinor.
\begin{equation}\label{eq:BPSODE}
    0 = \partial_w \eta +\f12 \left(\lambda_1^{\prime}+\lambda_2^{\prime}\right)\eta + \f m4\e^{-4(\lambda_1+\lambda_2)}g_1^{1/2}(\sigma^1\eta) + \f12 \left(\e^{-2\lambda_1}A_z^{\prime(1)}+\e^{-2\lambda_2}A_z^{\prime(2)}\right)(i\sigma^2\eta) \, .
\end{equation}
In order to solve these equations, note that the general form of the algebraic equations is given by,
\begin{equation}
	\begin{aligned}
		M_i\eta =& \left( X_i^{(0)}1 + X_i^{(1)}\sigma^1 + X_i^{(2)}~i\sigma^2 +
		X_i^{(3)}\sigma^3 \right)\eta = 0
        \,.
	\end{aligned}
\end{equation}
We can then collectively write all the equations as,
\begin{equation}
    \begin{bmatrix}
	\begin{array}{c}
        M_1 \\
        M_2\\
        \vdots
    \end{array}
	\end{bmatrix}
	\begin{bmatrix}
		\eta_1 \\
		\eta_2
	\end{bmatrix}
	=
	\begin{bmatrix}
		0 \\
	    0 \\
		0 \\
		0 \\
		\vdots
	\end{bmatrix}\,.
\end{equation}
A necessary condition to solve this system of equations is to impose that all
$2\times 2$ minors have to vanish. Defining 
\begin{equation}
	v_i = \begin{pmatrix}
		X_i^{(0)} + X_i^{(3)} & X_i^{(1)} + X_i^{(2)}
	\end{pmatrix}
	\,,\qquad 
	w_i = \begin{pmatrix}
		X_i^{(1)} - X_i^{(2)} & X_i^{(0)} - X_i^{(3)}
	\end{pmatrix}\,,
\end{equation}
we therefore impose
\begin{equation}
	\cA_{ij} = \det \begin{pmatrix}
		v_i \\ w_j
	\end{pmatrix} = 0\,,\qquad
	\cB_{ij} = \det \begin{pmatrix}
		v_i \\ v_j
	\end{pmatrix} = 0 \,,\qquad 
    \cC_{ij} = \det \begin{pmatrix}
		w_i \\ w_j
	\end{pmatrix} = 0 \,.
\end{equation}
These equations are clearly not all independent but a maximal set of independent equations is given for example by
\begin{equation}
\begin{aligned}
    0=&\, \cA_{11} = \cA_{12} = \cA_{13} = \cA_{14} = \cA_{15} = \cA_{16} = \cB_{12} = \cC_{13} \,.
\end{aligned}
\end{equation}
Our reparameterisation of the solutions is redundant and we are free to fix $g_2$. Rearranging and taking linear combinations allows us to find all the other fields by solving six algebraic equations and two non-linear differential equations. We checked that any solution to these equations is also a solution to the equations of motion. 

It is straightforward to check that these equations are indeed solved by the background given in \eqref{eq:7dmetric}, \eqref{eq:7dfields} provided the functions $h_i(w)$ satisfy the system of non-linear ODEs \eqref{eq:ODEsys}. Having showed that this background indeed solves the BPS equations and equations of motion we can proceed to solve the ODE \eqref{eq:BPSODE} and find the following explicit expression for the two-dimensional spinor,
\begin{equation}
    \eta=\e^{iqz}\f{w^{1/20}}{\sqrt{2}H(w)^{1/5}}\begin{bmatrix}
        \sqrt{H(w)^{1/2}+2w^{3/2} }\\
        \sqrt{H(w)^{1/2}-2w^{3/2}}
    \end{bmatrix}\,.
\end{equation}
Note that, as a function of $h_i(w)$, this spinor is identical to the case without the additional scalar, see e.g. \cite{Ferrero:2021etw}. The functions $h_i(w)$, however, are much more intricate when one adds the extra scalar.

\section{Absence of spindle solutions with extra scalars}
\label{app:spindle}

In this appendix, we give a proof for the absence of spindle solutions with additional scalars, analogous to the disc solutions presented in the main text. The general solution with the scalars $Y_i$ turned, \eqref{eq:7dmetric}-\eqref{eq:7dfields}, is completely determined in terms of the two functions $h_1(w)$ and $h_2(w)$. In order for these fields to solve the equations of motion and BPS equations, the functions $h_i(w)$ have to solve the system of ODEs \eqref{eq:ODEsys}, which we repeat here for convenience,
\begin{equation}
    f(w)\Big( h_i^\prime(w) -w h_i^{\prime\prime}(w) \Big) = w\f{H(w)}{h_i(w)}\left( \f14 h_i^\prime(w)^2-w^2 \right)\,.
\end{equation}
Solving this system of ODEs is prohibitively hard, and we cannot find a general solution. To go beyond the standard spindle/disc solution \eqref{eq:originalSol}, note that we are looking for functions $h_i$ such that $f$ has at least two real roots and is positive in the interval bounded by them. Therefore, we can make progress by solving the ODEs perturbatively around a root $w_0$ of $f$. To analyse these equations it will be useful to change coordinates to $W=w^2$ such that the ODEs take the form,
\begin{equation}\label{eq:ODEsW}
    4f(W)h_i^{\prime\prime}(W)+\f{H(W)}{h_i(W)}\left( h_i^{\prime}(W)^2-1 \right) = 0\,,
\end{equation}
where $f(w) = f(W)$ and the prime is now understood as a derivative with respect to $W$. Around a root $W=W_0$ of $f$ the ODEs reduce to $h_i^{\prime 2}(W) =1$, where only the positive sign makes sense given the expressions for the scalars $Y_i$, \eqref{eq:7dfields}. We expand the functions $h_i$ as follows
\begin{equation}\label{eq:expansion}
    h_i(W) = c_i + (W-W_0) + (W-W_0)^{\alpha_i} \sum_{k=1}^\infty a_k^{(i)} (W-W_0)^k \,,
\end{equation}
where the leading exponents $\alpha_i \in \bN_{>0}$. Demanding $h_i^{\prime}(W_0)=1$ fixes the linear term, while requiring that $f(W_0)=0$ fixes $c_2 = -4W_0^{3/2}/c_1$. Substituting these expansions in \eqref{eq:ODEsW}, for each pair of integers $\alpha_i$, the leading order determines the coefficients $c_1$ as well as the position of the root $W_0$ to be
\begin{equation}
    c_1 = \f{54\alpha_1\alpha_2^2}{(2+\alpha_1+\alpha_2)^3}\,,\qquad W_0 = \f{81\alpha_1^2\alpha_2^2}{(2+\alpha_1+\alpha_2)^4}\,.
\end{equation}
At the root the derivative of $f$ is given by
\begin{equation}
    f'(W_0) = -\f{27\alpha_1\alpha_2}{(2+\alpha_1+\alpha_2)^2} < 0\,,
\end{equation}
and therefore we see that when for both functions $h_i(w)$ contain higher order terms it is impossible to find solutions where the function $f$ has multiple roots and is positive in between. 

\begin{figure}[!ht]
    \centering
    \includegraphics[width=0.5\textwidth]{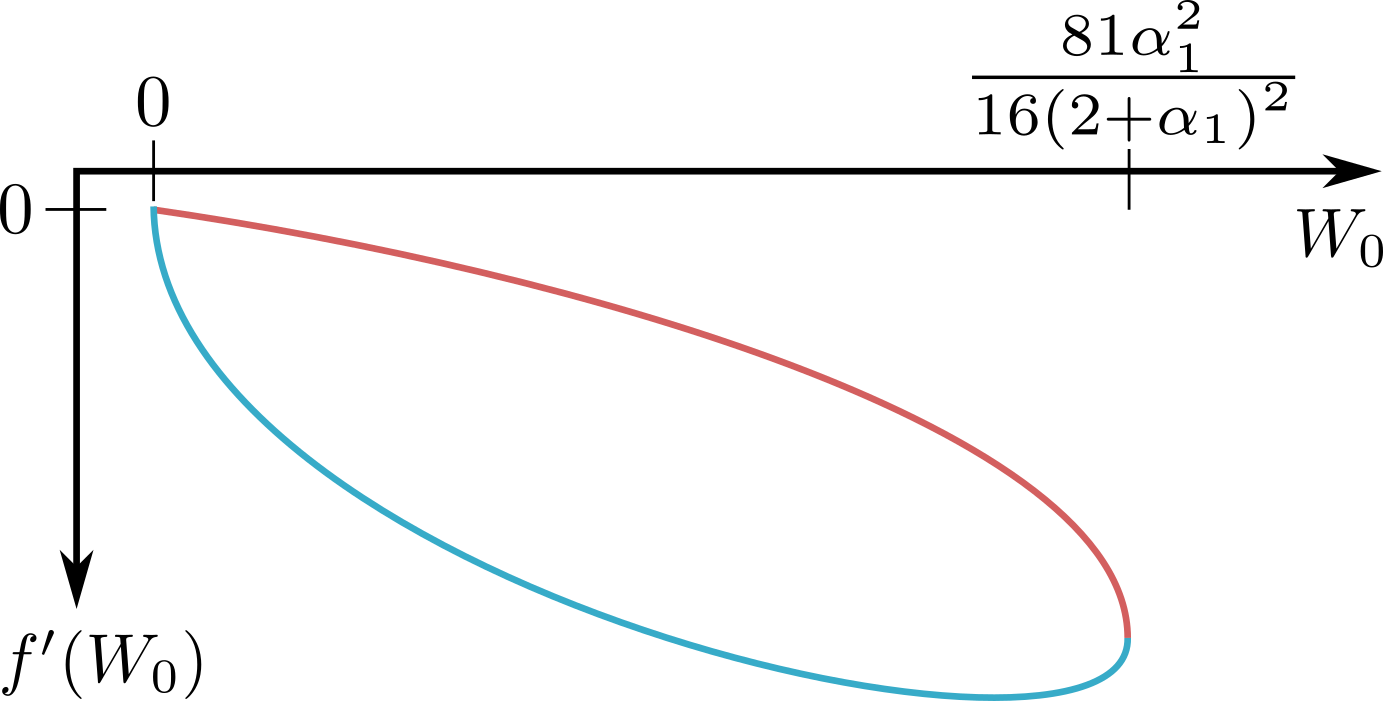}
    \caption{A plot of the derivative of $f$ at a root $W_0$ for generic values of $\alpha_1$. The red (blue) line corresponds to the plus (minus) sign in the expression for $f^\prime(W_0)$. Varying the value of $\alpha_1$ simply rescales the plot.}
    \label{fig:plotdf}
\end{figure}

In the above we demanded both functions to have higher order terms, however, we can still find spindle solutions when one of the functions, say $h_2$, is linear, with $c_2 \neq W_0$,
\begin{equation}
    h_2(W) = c_2 + (W-W_0)\,.
\end{equation}
Proceeding analogously as above we find that $c_i$ are completely fixed in terms of $\alpha_1$ and $W_0$,
\begin{equation}
    c_2 = -\f{4W_0^{3/2}}{c_1}\,,\qquad c_1 = 3W_0^{1/2}\pm \f{W_0^{1/2}}{\alpha_1^{1/2}}\left(9\alpha_1-4W_0^{1/2}\left(2+\alpha_1\right) \right)^{1/2}\,.
\end{equation}
In order to find real solutions we need to constrain the allowed range of $W_0$ to be
\begin{equation}
    W_0 \in \left[0 \,, \,\f{81\alpha_1^2}{16(2+\alpha_1)^2}\right]\,.
\end{equation}
Substituting the expansion in the derivative of $f$ at $W_0$ gives, 
\begin{equation}
    f'(W_0) = - \f{2W_0}{3 \alpha_1\pm \alpha_1^{1/2}\left(9\alpha_1-4W_0^{1/2}(2+\alpha_1)\right)^{1/2}} \leq 0\,.
\end{equation}
Within the allowed range for $W_0$ this function is always negative, see Figure \ref{fig:plotdf}, meaning that again we cannot find a solution with two nonzero roots with this ansatz for the expansion. Indeed, this is reflected in the observation that when we find numerical solutions to the system of ODEs, we always encounter a singularity when approaching a second root, indicating that at this point the ODEs cannot be solved.

The only way to avoid this is to set $q_2=0$, which in turn fixes one root to lie at $W=0$. This case is described in detail in the main text. The observation that for the spindle solutions the $\UU(1)^2$ isometry of the internal space cannot be explicitly broken by adding such scalars suggests that, in this case, the isometries do correspond to honest symmetries of their holographically dual SCFTs. This is in line with the analysis of \cite{BCtoappear}.

\section{Uplift formulae}
\label{app:Uplift}

In this work, we constructed various solutions of maximal $\SO(5)$ gauged supergravity. In order to uplift these solutions to eleven-dimensional supergravity solutions, we use the uplift formulae of \cite{Nastase:1999cb,Nastase:1999kf}. In terms of the seven-dimensional fields, the eleven-dimensional metric is given by,
\begin{equation}
    \ds_{11}^2 = \Delta^{1/3}\Big[\ds_7^2 +  \Delta^{-1} T^{-1}_{ij} \DD\mu^i \DD\mu^j\Big]\,,    
\end{equation}
where $\mu^i$, $i=1,\dots,5$ are embedding coordinates on $S^4$ satisfying 
\begin{equation}
    \sum_{i=1}^{5}(\mu^i)^2=1\,.
\end{equation} 
For our purposes a convenient parameterisation of the embedding coordinates is,
\begin{equation}
    \begin{aligned}
        \mu^1 =& \,\sqrt{1-\mu^2}\cos\phi \,,\quad    &   \mu^2 =&\, \sqrt{1-\mu^2}\sin\phi\,,\\
        \mu^3 =& \,\mu\cos\theta \,,\quad             &   \mu^4 =& \,\mu\sin\theta\cos\psi\,,\\
        \mu^5 =& \,\mu\sin\theta\sin\psi\,.
    \end{aligned}
\end{equation}
The function $\Delta$ is defined in terms of the scalar matrix $T_{ij}$ as,
\begin{equation}
    \Delta = T_{ij}\mu^i\mu^j\,,
\end{equation}
whilst the one-forms $\DD\mu^i$ are defined to be,
\begin{equation}
    \DD\mu^i = \dd \mu^i + g A^{ij}\mu_j\,,    
\end{equation}
where $A^{ij}$ are the SO$(5)$ gauge fields and $g$ the gauged supergravity coupling constant. This completely fixes the uplift of the metric and all that remains is to specify the form of the four-form flux of eleven-dimensional supergravity. This in turn is given by
\begin{equation}
\begin{aligned}\label{eq:G4uplift}
        G_4=& \frac{1}{4! g^3 \Delta^2}\,\epsilon_{i_1\dots i_5}\bigg[ -U \mu^{i_1}\DD\mu^{i_2}\wedge \DD\mu^{i_3}\wedge \DD\mu^{i_4}\wedge \DD\mu^{i_5}\\
        &\hspace{25.5mm}+4\mu^{j_1}\mu^{j_2} T^{i_1 j_1}\DD T^{i_2j_2} \wedge \DD\mu^{i_3}\wedge \DD\mu^{i_4}\wedge \DD\mu^{i_5}\\
        &\hspace{25.5mm}+6g\Delta   F^{i_1 i_2}\wedge \DD\mu^{i_3}\wedge \DD\mu^{i_4}\, T^{i_5 j} \mu^{j} \bigg]\\
        &- T_{ij} \star_7 S^{i} \mu^j+\frac{1}{g} S^{i} \wedge \DD\mu^i\, ,
\end{aligned}
\end{equation}
where $\star_7$ is the seven-dimensional Hodge star operator, and we defined
\begin{equation}
    U=2 T_{ij}T_{jk}\mu^{i}\mu^{j}-\Delta T_{ii}\, .
\end{equation}
When considering seven-dimensional $\AdS_5$ solutions, it is sufficient to restrict to a further $\UU(1)^{2}$ sub-truncation of the full seven-dimensional $\SO(5)$ gauged supergravity. In this sub-truncation, the only non-trivial $\SO(5)$ gauge fields are,
\begin{equation}
    A^{12} = A^{(1)}\, ,\qquad A^{34} = A^{(2)}\,.
\end{equation}
Demanding $\cN=2$ supersymmetry furthermore fixes $A^{(2)}$ to be zero. Furthermore, as the presence of a three-form is incompatible with the isometries of an $\AdS_5$ background all three-form gauge fields necessarily vanish in this sub-truncation.

When considering the more general solutions obtained through our consistent truncation to 5d we should relax some of these restrictions. Indeed, in this case we are forced to turn on a set of non-vanishing $\SU(2)$ gauge fields on top of the $\UU(1)$ gauge field $A^{(1)}$. Additionally, generically the three-form gauge fields $S^{I}$ will also be turned on. Indeed, from the branching rule,
\begin{equation}
    \SO(5) \rightarrow \SU(2)\times \UU(1)\quad :\quad \mathbf{5} \rightarrow \mathbf{3}_0 \oplus \mathbf{1}_2 \oplus \mathbf{1}_{-2}\,,
\end{equation}
we see that in the 5d theory these three-forms decompose as a neutral $\SU(2)$ triplet and two charged three-forms in line with the expected field content from 5d gauged supergravity where the $\mathbf{3}_0$ gives rise to the $\SU(2)$ gauge fields while the $\mathbf{1}_{\pm2}$ gives rise to the charged two form gauge fields.

\section{Consistent truncations for \texorpdfstring{$\mathcal{N}=2$ $\AdS_5$}{N=2 AdS5}  solutions}
\label{app:LLM}

In this appendix we spell out additional details regarding general $\cN=2$ preserving $\AdS_5$ solutions. In this paper we consider such solutions form various vantage points, we construct seven-dimensional $\AdS_5$ solutions, consider a truncation to five dimensions and along the way embed both types of solutions in the most general eleven-dimensional set-up \cite{Lin:2004nb}. The seven-dimensional solutions were constructed in maximal $\SO(5)$ gauged supergravity as discussed in detail in Appendix \ref{app:7dSUGRA}. The five-dimensional solutions on the other hand are constructed in 5d $\SU(2)\times \UU(1)$ Romans' gauged supergravity. We start this appendix by reviewing and introducing the most important aspects of both Romans' supergravity as well as the general eleven-dimension solution of \cite{Lin:2004nb}. After that we provide details omitted in the main text regarding the consistent truncation from LLM to Romans supergravity as constructed in \cite{Gauntlett:2007sm} as well as technical details for obtaining our novel consistent truncation from seven to five dimensions. 

\subsection{Romans' 5d \texorpdfstring{$\SU(2)\times\UU(1)$}{SU(2)xU(1)} gauged supergravity}
\label{app:5dSUGRA}

We start by giving some details on Romans' supergravity. The goal of this subsection is merely to set our conventions in order to resolve possible confusion. We therefore content ourselves with giving the field content and Lagrangian and refrain from giving explicitly the equations of motion and BPS equations. All the solutions constructed in this work can be embedded in seven- or eleven-dimensional supergravity and we checked that indeed they solve the five-, seven- and eleven-dimensional equations of motion and BPS equations. 

The bosonic field content of Romans' $\SU(2)\times\UU(1)$ gauged supergravity \cite{Romans:1985ps} consists of the metric, a real scalar field $X$, a collection of $\UU(1)\times\SU(2)$ gauge fields $\mathcal{B}$ and $\mathcal{A}^{a}$, with $a\in\{1,2,3\}$ and a complex two-form $\mathcal{C}$ whose real and imaginary part form a charged doublet with respect to the $\UU(1)$ gauge field. The Lagrangian for this theory is given by
\begin{equation}
    \begin{aligned}
    \mathcal{L}=&R\star 1-\frac{3}{X^2}\dd X\wedge \star \dd X -\frac{X^4}{2}\cG_2\wedge \star \cG_2-\frac{1}{2X^2} \Big( \cF^i_2\wedge \star \cF^i_2+\bar{\mathcal{C}}\wedge \star \cC\Big)\nonumber\\
    &-\ii \,\cC\wedge \bar{\cF_3}-\frac{1}{2} \cF^a_2\wedge \cF^a_2\wedge \mathcal{B}+ (X^2+2 X^{-1})\star 1\, ,
    \end{aligned}
\end{equation}
where with respect to \cite{Romans:1985ps} we set $g_1 = -1$ and $g_2 = -\sqrt{2}$. The field strengths appearing above are defined as
\begin{equation}
    \cG_2=\dd \cB\,, \qquad \cF^{a}_2\equiv\widetilde{\mathcal{D}}\mathcal{A}^a=\dd \cA^a -\frac{1}{2\sqrt{2}}\epsilon_{abc}\cA^b\wedge \cA^c \,,\qquad \cF_3= \dd \cC+\f{\ii}{2} \cB\wedge \cC\, .    
\end{equation}
The kinetic term for the scalar field may be written in the canonical form by introducing a dilaton field $\phi$ defined through $X=\e^{-\tfrac{1}{\sqrt{6}}\phi}$. One can further truncated this theory to 5d Einstein--Maxwell, a.k.a. minimal, supergravity by letting $\phi=\cF^1_2=\cF^2_2=\cF_3=0$ vanish and identifying $\cF^3_2=\sqrt{2}\cG_2 = \sqrt{\f13}\cF$ and $\cB= \sqrt{\f13}\cA$. The resulting Lagrangian is
\begin{equation}
    \mathcal{L}=(R-3)\star 1 -\frac{1}{2}\cF\wedge \star \cF-\frac{1}{3\sqrt 3} \cF\wedge \cF\wedge \cA\,.
\end{equation}
%

\subsection{\texorpdfstring{$\mathcal{N}=2$ $\AdS_5$}{N=2 AdS5} solutions of 11d supergravity}
\label{app:LLMform}

Having discussed the 5d Romans' supergravity we move on to discussing the most general $\cN=2$ preserving $\AdS_5$ background of eleven-dimensional supergravity as introduced in \cite{Lin:2004nb}. 

The metric and four-form flux for this background are given by
\begingroup
\allowdisplaybreaks
\begin{align}\label{eq:LLM}
    \ds^2 =&~ \e^{2\lambda}\left( 4\ds^2_{\AdS_5} + y^2 \e^{-6\lambda} \ds^2_{S^2} + \ds^2_4 \right)\,,\nn\\
    ds^2_4 =&~ 4\left( 1- y^2 \e^{-6\lambda} \right) \left( \dd\chi + v \right)^2 + \f{\e^{-6\lambda}}{1-y^2 \e^{-6\lambda}}\left( \dd y^2 + \e^D( \dd x_1^2 + \dd x_2^2 ) \right)\,,\nn\\
    G_4 =&~ F \wedge \dvol(S^2)\,,\nn\\
    \e^{-6\lambda} =& - \f{\partial_y D}{y(1-y\partial_y D)}\,,\\
    v =&~ \f12 \epsilon_{ij}\partial_{j} D\dd x^{i}\,,\nn\\
    F =& ~\dd \big(4y^3 \e^{-6\lambda}  (\dd\chi + v)\big)  + \dd \hat{B}\,,\nn\\
    \dd\hat{B}=&~ 4 y \dd v-2 \partial_y\me^{D}\dd x_1\wedge \dd x_2\,.\nn
\end{align}
\endgroup
Note that this solution is completely determined by a single potential $D$ which satisfies the $\SU(\infty)$ Toda equation,
\begin{equation}
    \square D + \partial_y^2 \e^D = \left( \partial^2_{x^1} +\partial^2_{x^2}\right) D +  \partial_y^2 \e^D = 0\,. 
\end{equation}
In this way any solution of the above Toda equations can be mapped to an $\cN=2$ preserving $\AdS_5$ solution in eleven-dimensional supergravity. 

In section \ref{sec:Analysis}, in order to compute the conformal dimensions of the BPS particles corresponding to wrapped probe M2-branes we need to formulate a calibration condition in order to ensure supersymmetry of said operators. In \cite{Gauntlett:2006ai} a generalised calibration was described for the most general $\cN=1$ $\AdS_5$ solution of M-theory \cite{Gauntlett:2004zh}. Specifying to the LLM system, the calibration 2-form $\cX$ was given in \cite{Bah:2021hei}. We refer the reader to the above mentioned references for more details on its construction. In terms of the LLM fields it takes the form,
\begin{equation}\label{eq:calX}
\begin{aligned}
    \cX=&\, y^3 \e^{-9\lambda} \dvol(S^2) + y \e^{-3\lambda}(1-y^2\e^{-6\lambda}) \dd \tau \wedge (\dd\chi+v) \\
    &\,- \tau \e^{-3\lambda} (\dd\chi+v)\wedge \dd y + \f{\tau y \e^{-9\lambda}\e^D}{1-y^2 \e^{-6\lambda}}\dd x_1\wedge \dd x_1\,,
\end{aligned}
\end{equation}
where the metric on the round two-sphere is given by
\begin{equation}\label{eq:S2met}
    \ds^2_{S^2} = \f{\dd \tau^2}{1-\tau^2} + (1-\tau^2) \dd\varphi^2\,, \qquad\qquad\text{where }\quad \tau\in [ -1,1]\,.
\end{equation}
%

\subsection{Consistent truncation of LLM to 5d gauged supergravity}
\label{app:5dLLM}

Having introduced both Romans' supergravity in five dimensions as well as the eleven-dimensional LLM geometry we are ready to discuss the consistent truncation of the latter to the former \cite{Gauntlett:2007sm}.\footnote{Note that our notation differs slightly from that in \cite{Gauntlett:2007sm}. The coordinates and functions here are related to theirs as follows, 
\begin{equation}
z_{\rm GV} = y^2 \mathrm{e}^{-6\lambda}, \qquad 
\lambda_{\rm GV} = \e^{-2\lambda}, \qquad \rho_{\rm GV} = y\,.
\end{equation}
In addition, we fixed the overall constant $m=\f12$.
} Here we give all the necessary formulae needed in the main text, for more details we refer the reader to the original reference.

In order to truncate the LLM geometry to five dimension, we rewrite the metric as follows,
\begin{align}
\dd s^2_{11} &= \Big(\frac{\Omega}{X}\Big)^{1/3}\me^{2\lambda}\bigg[4\dd s^2_5 +\frac{X \me^{-6 \lambda}}{1-y^2 \me^{-6 \lambda}}\Big(\dd y^2+\me^{D}\big(\dd x_1^2+\dd x_2^2\big)\Big)\nonumber\\
&+\frac{4 X^2(1-y^2\me^{-6 \lambda})}{\Omega}(\dd\chi+v+\tfrac{1}{2} \mathcal{B})^2 +\frac{y^2\me^{-6 \lambda}}{X \Omega}\widetilde{\mathcal{D}}\tilde{\mu}^a \widetilde{\mathcal{D}}\tilde{\mu}^a \bigg]\, ,
\end{align}
where we defined the function
\begin{equation}
    \Omega= X y^2\me^{-6 \lambda}+ X^{-2}(1-y^2\me^{-6 \lambda})\, .
\end{equation}
The sphere is defined in terms of the embedding coordinates $\tilde\mu^a$ which satisfy $\sum_{a=1}^{3} (\tilde\mu^a)^2 = 1$ The gauged one-forms appearing in the metric are defined as
\begin{equation}
    \DDt\tilde{\mu}^a=\dd\tilde{\mu}^a +\frac{1}{\sqrt{2}}\epsilon_{abc}\tilde{\mu}^{b}\mathcal{A}^{c}\, .    
\end{equation}
Rewriting the four-form flux in terms of the five-dimensional fields is slightly more complicated, with the resulting expression for $G_4$ being
\begin{align}\label{eq:G4form}
    G_4 =&\, \Tilde{G}_4 + \cG_2 \wedge \beta_2 + \cF^a_2 \wedge \beta^a_2 + \star_5 \cF^a_2 \wedge \beta^a_1 + (\mathcal{C} \wedge \alpha_2 + \cF_2 \wedge \alpha_1 + c.c.)\,,\\
    \begin{split}
        \Tilde{G}_4 =&\,- \frac{1}{2} \epsilon_{abc} \tilde{\mu}^a \widetilde{\mathcal{D}}\tilde{\mu}^b \wedge \widetilde{\mathcal{D}}\tilde{\mu}^c \wedge \bigg[ \dd \Big(\frac{\me^{6\lambda}(-\partial_y D)}{4 X^2 \Omega} \Big) \wedge \frac{1}{2 \me^\lambda \sqrt{1-y^2\me^{-6 \lambda}}} \hat{e}^3\\
        &+ \frac{\me^{6\lambda}(-\partial_y D)}{4 X^2 \Omega} \wedge \dd \left( \frac{1}{2 \me^\lambda \sqrt{1-y^2\me^{-6 \lambda}}} e^3 \right) + \Big( \frac{2y}{\me^{2\lambda}} e^1 \wedge e^2 + 2 \me^{\lambda} \hat{e}^3 \wedge e^4 \Big) \bigg]\,.
    \end{split}
\end{align}
In these expressions, $\star_5$ denotes the Hodge dual with respect to the five-dimensional metric $\ds_5^2$ and $c.c.$ denotes the complex conjugated expression. The internal components of the four-form flux are in turn given by
\begin{equation}
\begin{aligned}\label{eq:truncforms}
\beta_2 & = \frac{Xy^3\e^{-6\lambda}}{2\Omega} \epsilon_{abc} \,\tilde\mu^a \,\DDt \tilde\mu^b \wedge \DDt \tilde\mu^c\,, \\
\beta_2^a & =\frac{1}{\sqrt{2}}\left[\f{y\,\sqrt{1-y^2\e^{-6\lambda}}}{X^{2}\Omega}  \e^{-\lambda}\, \DDt \tilde\mu^a \wedge \hat{e}^3- \tilde\mu^a\left(y\,\e^{-2\lambda}\, e^{12}+\e^{\lambda}\, \hat{e}^{34}\right)\right]\,, \\
\beta_1^a & =-\frac{1}{\sqrt{2}X^2}\left(\tilde{\mu}^a\, \dd y+y \,\DDt \tilde{\mu}^a\right)\,, \\
\alpha_1 & =\frac{1}{\sqrt{2}} \e^{2\lambda}\,\sqrt{1-y^2\e^{-6\lambda}}\left(e^1-i e^2\right), \\
\alpha_2 & =\frac{1}{2 \sqrt{2}}\left(e^1-i e^2\right) \wedge\left(y\,\e^{-2\lambda} e^4+i \e^{\lambda} \,\hat{e}^3\right) \,.
\end{aligned}
\end{equation}
In the expression above, we defined $e^{ab}=e^a\wedge e^b$ (and similarly $\hat e^{3a}=\hat{e}^3\wedge e^a$) where the vielbein $e^a$ are 
\begin{equation}
    \begin{split}
        e^1 &= \f{\e^{-2\lambda}\e^{D/2}}{\sqrt{1-y^2\e^{-6\lambda}}}\big(\sin \chi\dd x_1+\cos\chi \dd x_2\big) \,, \\
        e^2 &= \f{\e^{-2\lambda}\e^{D/2}}{\sqrt{1-y^2\e^{-6\lambda}}} \big(-\cos\chi \dd x_1+\sin\chi\dd x_2\big)\,,\\
        e^3 &= 2\e^{\lambda}\sqrt{1-y^2\e^{-6\lambda}}(\dd\chi + v) \,,  \\
        e^4 &= \f{\e^{-2\lambda}\dd y}{\sqrt{1-y^2\e^{-6\lambda}}} \,.
    \end{split}
\end{equation}
Finally $\hat{e}^3$ denotes the gauging of the $e^3$ vielbein by the gauge field $\mathcal{B}$ via $\dd \chi\rightarrow \dd\chi+\tfrac{1}{2}\mathcal{B}$.
\begin{equation}
    \hat{e}^3 = 2\e^{\lambda}\sqrt{1-y^2\e^{-6\lambda}}\Big(\dd\chi + v + \frac{1}{2} \mathcal{B}\Big)\, .
\end{equation}
This concludes all the necessary expressions specifying the consistent truncation of the LLM geometry to five-dimensional Romans' supergravity.

\subsection{Consistent truncation of 7d to 5d gauged supergravity}
\label{app:5d7d}

Finally, all the ingredients are in place to discuss the truncation of seven-dimensional maximal $\SO(5)$ gauged supergravity to five-dimensional Romans' supergravity. In the main text we discuss how to extract the metric, scalars and one-form gauge fields as these can straightforwardly be extracted from the eleven-dimensional metric. In order complete the truncation of the 7d theory we are left with determining the correct form of the three-forms $S^i$. This, however, is more subtle, since they appear only in the flux $G_4$ and therefore we need to carefully study the flux to extract these fields.

To do so let us start by defining the gauged volume,
\begin{equation}
    \text{vol} (S^2)^g = 
    \frac{1}{2} \epsilon_{abc} \tilde{\mu}^a D\tilde{\mu}^b \wedge D\tilde{\mu}^c\, ,
\end{equation}
which reduces to the volume of the unit radius two-sphere when the $\SU(2)$ gauge fields are set to vanish. Next, in order to efficiently compare the flux we rewrite it to take the form \eqref{eq:G4form}. The $\beta$-forms appearing in \eqref{eq:truncforms} are
\begin{equation}
\begin{split}
    \beta_2=&\,\frac{y^3}{ \tilde{\Omega}}\dvol(S^2)^g\, ,\\
    \beta_1^a=&\,\frac{1}{4 \sqrt{2} X^2}\big(w \tilde{\mathcal{D}}\tilde{\mu}^{a+2}+\tilde{\mu}^{a+2}\dd w\Big)\, ,
    \end{split}
\end{equation}
while the $\alpha$'s are
\begin{equation}
\begin{split}
    \alpha_1=&\,\e^{-\ii \chi}\bigg[ \frac{\sqrt{f(w)}}{2 \sqrt{2} w}\big(\ii \e^{Y_1(w)/2}\tilde{\mathcal{D}}\tilde{\mu}^1+\e^{-Y_1(w)/2}\tilde{\mathcal{D}}\tilde{\mu}^2\big)\\
    &+\frac{1}{2\sqrt{2}}\big(\ii \e^{-Y_1(w)/2}\tilde{\mu}^1+\e^{Y_1(w)/2}\tilde{\mu}^2\big)\Big[\frac{w^2}{4\sqrt{f(w)}}\dd w-\ii \frac{\sqrt{f(w)}}{w}A^{12}\Big]\bigg]\, ,\\
\alpha_2=&\,\frac{1}{2\sqrt{2}}\e^{-\ii \chi}\bigg[\Big(\e^{Y_1(w)/2}\mathcal{D}\tilde{\mu}^1-\ii \e^{-Y_1(w)/2}\tilde{\mathcal{D}}\tilde{\mu}^2\Big)\wedge \Big(\frac{\ii w}{2\sqrt{f(w)}}\dd w-\frac{\sqrt{f(w)}}{w} D\chi\Big)\\
&+\frac{\ii w^2 \sqrt{f(w)}}{4 (f(w)+w^3 X^3)}\big(\ii \e^{-Y_1(w)/2}\mu^1+\e^{Y_1(w)/2}\mu^2\big) \dd w\wedge D\chi\bigg]\, ,
\end{split}
\end{equation}
where $D\chi$ is defined in equation \eqref{eq:fibrationcoord}.

Similarly, we can extract the external part of the four-form flux to find,
\begin{align}
    \tilde{G}_4 =&\,-2 \dvol (S^2)^g \wedge \bigg[ \dd \Big[ 2 y\Big(1-\frac{y^2}{\tilde{\Omega}}\Big)\Big] (D\chi + v ) +2 y\Big(1-\frac{y^2}{\tilde{\Omega}}\Big)\dd v \nonumber\\
    &- 2 \dd y \wedge (D\chi + v) - \partial_y D \, \dd x^1 \wedge \dd x^2 \bigg] \, .
\end{align}
Note that $\tilde{\Omega}$ was given in \eqref{eq:Omt} and we have refrained from giving the explicit form of $\beta_2^a$ since it is particularly unwieldy and we were able to explicitly extract this term out of the flux from the 7d uplift below.

With these expressions at hand we can now compare these terms with the flux arising from uplifting a seven-dimensional solution of maximal gauge supergravity, see \eqref{eq:G4uplift}. Carefully rewriting the uplifted $G_4$ for our 7d truncation results in the following terms
\begin{equation}
    \begin{split}
        G_4^{(1)}=&\,\frac{\mu^2 U}{\Delta^2}\dvol(S^2)^g \wedge \dd \mu\wedge \Big(\dd\phi- A^{12}\Big)\, ,\\
        G_4^{(2)}=&\,-\frac{2(1-\mu^2)\mu^2\hat{X}_1^2}{\Delta^2}\dvol(S^2)^g\wedge \dd \mu\wedge\Big(  \cos\phi\sin\phi \dd Y_1 \\
        &\,+ \sinh(Y_1(w))\big(\e^{-Y_1(w)}\cos^2\phi-\e^{Y_1(w)}\sin^2\phi\big)A^{12}\Big)\, ,\\
        &\,+\mu^3\bigg(\hat{X}_2^2\Big(\tilde{\mu}^1 \dd\big(\e^{-Y_1(w)}\hat{X}_1\hat{X}_2^{-1}\big)\wedge \tilde{\mathcal{D}}\tilde{\mu}^2-\tilde{\mu}^2 \dd\big(\e^{Y_1(w)}\hat{X}_1\hat{X}_2^{-1}\big)\wedge \tilde{\mathcal{D}}\tilde{\mu}^1\Big)\\
        &\,+2\sinh\big(Y_1(w)\big)\hat{X}_1\hat{X}_2 A^{12}\wedge \big(\tilde{\mu}^2\tilde{\mathcal{D}}\tilde{\mu}^2-\tilde{\mu}^1\tilde{\mathcal{D}}\tilde{\mu}^2\big)\bigg)\wedge\dvol(S^2)^g \, ,\\
        G_4^{(3)}=&\,-\beta_2\wedge \dd\mathcal{B}-\beta^a_2\wedge\mathcal{F}^a\\
        &\,-\frac{\hat{X}_1^{5/3} \mu^3}{16 X w^2\tilde{\Omega}}\big(3 w f(w)\dd X+X(w^3 X^3 +4 f(w)-w f'(w))\dd w\big)\wedge D\chi\\
        &\,- \frac{w f(w)}{2\sqrt{2}(f(w)+w^3 X^3)}\mathcal{F}^a\wedge D\chi\wedge\tilde{\mathcal{D}}\tilde{\mu}^a+2+\frac{1}{2\sqrt{2}}\tilde{\mu}^{a+2}\dd w\wedge D\chi\wedge \mathcal{F}^a\, .
    \end{split}
\end{equation}
The labels in superscript indicate the origin of the respective term in the uplift of 7d maximal gauged supergravity and corresponds to the line number in equation \eqref{eq:G4uplift}. 
After all this preparatory work we can now simply compare the terms of the two truncations to obtain the resulting expressions for the three-forms  \eqref{eq:threeforms} presented in the main text. By construction, the truncation of the 7d theory on the disc is consistent.


\bibliographystyle{JHEP}
\bibliography{sample}

\providecommand{\href}[2]{#2}\begingroup\raggedright\begin{thebibliography}{10}

\bibitem{Argyres:1995jj}
P.~C. Argyres and M.~R. Douglas, {\it {New phenomena in SU(3) supersymmetric
  gauge theory}},  {\em Nucl. Phys. B} {\bf 448} (1995) 93--126,
  [\href{http://arxiv.org/abs/hep-th/9505062}{{\tt hep-th/9505062}}].

\bibitem{Xie:2012hs}
D.~Xie, {\it {General Argyres-Douglas Theory}},  {\em JHEP} {\bf 01} (2013)
  100, [\href{http://arxiv.org/abs/1204.2270}{{\tt arXiv:1204.2270}}].

\bibitem{Argyres:1995xn}
P.~C. Argyres, M.~R. Plesser, N.~Seiberg, and E.~Witten, {\it {New N=2
  superconformal field theories in four-dimensions}},  {\em Nucl. Phys. B} {\bf
  461} (1996) 71--84, [\href{http://arxiv.org/abs/hep-th/9511154}{{\tt
  hep-th/9511154}}].

\bibitem{Eguchi:1996vu}
T.~Eguchi, K.~Hori, K.~Ito, and S.-K. Yang, {\it {Study of N=2 superconformal
  field theories in four-dimensions}},  {\em Nucl. Phys. B} {\bf 471} (1996)
  430--444, [\href{http://arxiv.org/abs/hep-th/9603002}{{\tt hep-th/9603002}}].

\bibitem{Gaiotto:2009hg}
D.~Gaiotto, G.~W. Moore, and A.~Neitzke, {\it {Wall-crossing, Hitchin Systems,
  and the WKB Approximation}},  \href{http://arxiv.org/abs/0907.3987}{{\tt
  arXiv:0907.3987}}.

\bibitem{Bonelli:2011aa}
G.~Bonelli, K.~Maruyoshi, and A.~Tanzini, {\it {Wild Quiver Gauge Theories}},
  {\em JHEP} {\bf 02} (2012) 031, [\href{http://arxiv.org/abs/1112.1691}{{\tt
  arXiv:1112.1691}}].

\bibitem{Wang:2015mra}
Y.~Wang and D.~Xie, {\it {Classification of Argyres-Douglas theories from M5
  branes}},  {\em Phys. Rev. D} {\bf 94} (2016), no.~6 065012,
  [\href{http://arxiv.org/abs/1509.00847}{{\tt arXiv:1509.00847}}].

\bibitem{Bah:2021mzw}
I.~Bah, F.~Bonetti, R.~Minasian, and E.~Nardoni, {\it {Holographic Duals of
  Argyres-Douglas Theories}},  {\em Phys. Rev. Lett.} {\bf 127} (2021), no.~21
  211601, [\href{http://arxiv.org/abs/2105.11567}{{\tt arXiv:2105.11567}}].

\bibitem{Bah:2021hei}
I.~Bah, F.~Bonetti, R.~Minasian, and E.~Nardoni, {\it {M5-brane sources,
  holography, and Argyres-Douglas theories}},  {\em JHEP} {\bf 11} (2021) 140,
  [\href{http://arxiv.org/abs/2106.01322}{{\tt arXiv:2106.01322}}].

\bibitem{Couzens:2022yjl}
C.~Couzens, H.~Kim, N.~Kim, and Y.~Lee, {\it {Holographic duals of M5-branes on
  an irregularly punctured sphere}},  {\em JHEP} {\bf 07} (2022) 102,
  [\href{http://arxiv.org/abs/2204.13537}{{\tt arXiv:2204.13537}}].

\bibitem{Bah:2022yjf}
I.~Bah, F.~Bonetti, E.~Nardoni, and T.~Waddleton, {\it {Aspects of irregular
  punctures via holography}},  {\em JHEP} {\bf 11} (2022) 131,
  [\href{http://arxiv.org/abs/2207.10094}{{\tt arXiv:2207.10094}}].

\bibitem{Maldacena:2000mw}
J.~M. Maldacena and C.~Nunez, {\it {Supergravity description of field theories
  on curved manifolds and a no go theorem}},  {\em Int. J. Mod. Phys.} {\bf
  A16} (2001) 822--855, [\href{http://arxiv.org/abs/hep-th/0007018}{{\tt
  hep-th/0007018}}]. [,182(2000)].

\bibitem{Bah:2011vv}
I.~Bah, C.~Beem, N.~Bobev, and B.~Wecht, {\it {AdS/CFT Dual Pairs from
  M5-Branes on Riemann Surfaces}},  {\em Phys. Rev.} {\bf D85} (2012) 121901,
  [\href{http://arxiv.org/abs/1112.5487}{{\tt arXiv:1112.5487}}].

\bibitem{Bah:2012dg}
I.~Bah, C.~Beem, N.~Bobev, and B.~Wecht, {\it {Four-Dimensional SCFTs from
  M5-Branes}},  {\em JHEP} {\bf 06} (2012) 005,
  [\href{http://arxiv.org/abs/1203.0303}{{\tt arXiv:1203.0303}}].

\bibitem{Bobev:2019ore}
N.~Bobev, P.~Bomans, and F.~F. Gautason, {\it {Wrapped Branes and Punctured
  Horizons}},  \href{http://arxiv.org/abs/1912.04779}{{\tt arXiv:1912.04779}}.

\bibitem{Ferrero:2020laf}
P.~Ferrero, J.~P. Gauntlett, J.~M. P\'erez Ipi\~na, D.~Martelli, and J.~Sparks,
  {\it {D3-Branes Wrapped on a Spindle}},  {\em Phys. Rev. Lett.} {\bf 126}
  (2021), no.~11 111601, [\href{http://arxiv.org/abs/2011.10579}{{\tt
  arXiv:2011.10579}}].

\bibitem{Ferrero:2020twa}
P.~Ferrero, J.~P. Gauntlett, J.~M.~P. Ipi\~na, D.~Martelli, and J.~Sparks, {\it
  {Accelerating black holes and spinning spindles}},  {\em Phys. Rev. D} {\bf
  104} (2021), no.~4 046007, [\href{http://arxiv.org/abs/2012.08530}{{\tt
  arXiv:2012.08530}}].

\bibitem{Hosseini:2021fge}
S.~M. Hosseini, K.~Hristov, and A.~Zaffaroni, {\it {Rotating multi-charge
  spindles and their microstates}},  {\em JHEP} {\bf 07} (2021) 182,
  [\href{http://arxiv.org/abs/2104.11249}{{\tt arXiv:2104.11249}}].

\bibitem{Boido:2021szx}
A.~Boido, J.~M.~P. Ipi\~na, and J.~Sparks, {\it {Twisted D3-brane and M5-brane
  compactifications from multi-charge spindles}},  {\em JHEP} {\bf 07} (2021)
  222, [\href{http://arxiv.org/abs/2104.13287}{{\tt arXiv:2104.13287}}].

\bibitem{Faedo:2021kur}
F.~Faedo, S.~Klemm, and A.~Vigan\`o, {\it {Supersymmetric black holes with
  spiky horizons}},  {\em JHEP} {\bf 09} (2021) 102,
  [\href{http://arxiv.org/abs/2105.02902}{{\tt arXiv:2105.02902}}].

\bibitem{Ferrero:2021wvk}
P.~Ferrero, J.~P. Gauntlett, D.~Martelli, and J.~Sparks, {\it {M5-branes
  wrapped on a spindle}},  {\em JHEP} {\bf 11} (2021) 002,
  [\href{http://arxiv.org/abs/2105.13344}{{\tt arXiv:2105.13344}}].

\bibitem{Cassani:2021dwa}
D.~Cassani, J.~P. Gauntlett, D.~Martelli, and J.~Sparks, {\it {Thermodynamics
  of accelerating and supersymmetric AdS4 black holes}},  {\em Phys. Rev. D}
  {\bf 104} (2021), no.~8 086005, [\href{http://arxiv.org/abs/2106.05571}{{\tt
  arXiv:2106.05571}}].

\bibitem{Ferrero:2021ovq}
P.~Ferrero, M.~Inglese, D.~Martelli, and J.~Sparks, {\it {Multicharge
  accelerating black holes and spinning spindles}},  {\em Phys. Rev. D} {\bf
  105} (2022), no.~12 126001, [\href{http://arxiv.org/abs/2109.14625}{{\tt
  arXiv:2109.14625}}].

\bibitem{Couzens:2021rlk}
C.~Couzens, K.~Stemerdink, and D.~van~de Heisteeg, {\it {M2-branes on discs and
  multi-charged spindles}},  {\em JHEP} {\bf 04} (2022) 107,
  [\href{http://arxiv.org/abs/2110.00571}{{\tt arXiv:2110.00571}}].

\bibitem{Faedo:2021nub}
F.~Faedo and D.~Martelli, {\it {D4-branes wrapped on a spindle}},  {\em JHEP}
  {\bf 02} (2022) 101, [\href{http://arxiv.org/abs/2111.13660}{{\tt
  arXiv:2111.13660}}].

\bibitem{Ferrero:2021etw}
P.~Ferrero, J.~P. Gauntlett, and J.~Sparks, {\it {Supersymmetric spindles}},
  {\em JHEP} {\bf 01} (2022) 102, [\href{http://arxiv.org/abs/2112.01543}{{\tt
  arXiv:2112.01543}}].

\bibitem{Couzens:2021cpk}
C.~Couzens, {\it {A tale of (M)2 twists}},  {\em JHEP} {\bf 03} (2022) 078,
  [\href{http://arxiv.org/abs/2112.04462}{{\tt arXiv:2112.04462}}].

\bibitem{Giri:2021xta}
S.~Giri, {\it {Black holes with spindles at the horizon}},  {\em JHEP} {\bf 06}
  (2022) 145, [\href{http://arxiv.org/abs/2112.04431}{{\tt arXiv:2112.04431}}].

\bibitem{Couzens:2022agr}
C.~Couzens, N.~T. Macpherson, and A.~Passias, {\it {On Type IIA AdS$_{3}$
  solutions and massive GK geometries}},  {\em JHEP} {\bf 08} (2022) 095,
  [\href{http://arxiv.org/abs/2203.09532}{{\tt arXiv:2203.09532}}].

\bibitem{Cheung:2022wpg}
K.~C.~M. Cheung and R.~Leung, {\it {Type IIA embeddings of D = 5 minimal gauged
  supergravity via non-Abelian T-duality}},  {\em JHEP} {\bf 06} (2022) 051,
  [\href{http://arxiv.org/abs/2203.15114}{{\tt arXiv:2203.15114}}].

\bibitem{Suh:2022olh}
M.~Suh, {\it {M5-branes and D4-branes wrapped on a direct product of spindle
  and Riemann surface}},  \href{http://arxiv.org/abs/2207.00034}{{\tt
  arXiv:2207.00034}}.

\bibitem{Arav:2022lzo}
I.~Arav, J.~P. Gauntlett, M.~M. Roberts, and C.~Rosen, {\it {Leigh-Strassler
  compactified on a spindle}},  {\em JHEP} {\bf 10} (2022) 067,
  [\href{http://arxiv.org/abs/2207.06427}{{\tt arXiv:2207.06427}}].

\bibitem{Couzens:2022yiv}
C.~Couzens and K.~Stemerdink, {\it {Universal spindles: D2's on $\Sigma$ and
  M5's on $\Sigma\times \mathbb{H}^3$}},
  \href{http://arxiv.org/abs/2207.06449}{{\tt arXiv:2207.06449}}.

\bibitem{Couzens:2022aki}
C.~Couzens, N.~T. Macpherson, and A.~Passias, {\it {A plethora of Type IIA
  embeddings for d = 5 minimal supergravity}},  {\em JHEP} {\bf 01} (2023) 047,
  [\href{http://arxiv.org/abs/2209.15540}{{\tt arXiv:2209.15540}}].

\bibitem{Boido:2022mbe}
A.~Boido, J.~P. Gauntlett, D.~Martelli, and J.~Sparks, {\it {Gravitational
  Blocks, Spindles and GK Geometry}},
  \href{http://arxiv.org/abs/2211.02662}{{\tt arXiv:2211.02662}}.

\bibitem{Suh:2022pkg}
M.~Suh, {\it {Spindle black holes from mass-deformed ABJM}},
  \href{http://arxiv.org/abs/2211.11782}{{\tt arXiv:2211.11782}}.

\bibitem{Suh:2023xse}
M.~Suh, {\it {Baryonic spindles from conifolds}},
  \href{http://arxiv.org/abs/2304.03308}{{\tt arXiv:2304.03308}}.

\bibitem{Amariti:2023mpg}
A.~Amariti, N.~Petri, and A.~Segati, {\it {T$^{1,1}$ truncation on the
  spindle}},  {\em JHEP} {\bf 07} (2023) 087,
  [\href{http://arxiv.org/abs/2304.03663}{{\tt arXiv:2304.03663}}].

\bibitem{Kim:2023ncn}
H.~Kim, N.~Kim, Y.~Lee, and A.~Poole, {\it {Thermodynamics of accelerating
  AdS$_4$ black holes from the covariant phase space}},
  \href{http://arxiv.org/abs/2306.16187}{{\tt arXiv:2306.16187}}.

\bibitem{Hristov:2023rel}
K.~Hristov and M.~Suh, {\it {Spindle black holes in AdS$_4 \times$SE$_7$}},
  \href{http://arxiv.org/abs/2307.10378}{{\tt arXiv:2307.10378}}.

\bibitem{Couzens:2021tnv}
C.~Couzens, N.~T. Macpherson, and A.~Passias, {\it {$ \mathcal{N} $ = (2, 2)
  AdS$_{3}$ from D3-branes wrapped on Riemann surfaces}},  {\em JHEP} {\bf 02}
  (2022) 189, [\href{http://arxiv.org/abs/2107.13562}{{\tt arXiv:2107.13562}}].

\bibitem{Suh:2021ifj}
M.~Suh, {\it {D3-branes and M5-branes wrapped on a topological disc}},  {\em
  JHEP} {\bf 03} (2022) 043, [\href{http://arxiv.org/abs/2108.01105}{{\tt
  arXiv:2108.01105}}].

\bibitem{Suh:2021aik}
M.~Suh, {\it {D4-branes wrapped on a topological disk}},  {\em JHEP} {\bf 06}
  (2023) 008, [\href{http://arxiv.org/abs/2108.08326}{{\tt arXiv:2108.08326}}].

\bibitem{Suh:2021hef}
M.~Suh, {\it {M2-branes wrapped on a topological disk}},  {\em JHEP} {\bf 09}
  (2022) 048, [\href{http://arxiv.org/abs/2109.13278}{{\tt arXiv:2109.13278}}].

\bibitem{Karndumri:2022wpu}
P.~Karndumri and P.~Nuchino, {\it {Five-branes wrapped on topological disks
  from 7D N=2 gauged supergravity}},  {\em Phys. Rev. D} {\bf 105} (2022),
  no.~6 066010, [\href{http://arxiv.org/abs/2201.05037}{{\tt
  arXiv:2201.05037}}].

\bibitem{Couzens:toappear}
C.~Couzens, M.~Jinwoo-Kang, C.~Lawrie, and Y.~Lee, {\it {Holographic duals of
  Higgsed $\mathcal{D}_p^{\,b}(BCD)$}},  {\em To appear} (2023).

\bibitem{Cheung:2022ilc}
K.~C.~M. Cheung, J.~H.~T. Fry, J.~P. Gauntlett, and J.~Sparks, {\it {M5-branes
  wrapped on four-dimensional orbifolds}},  {\em JHEP} {\bf 08} (2022) 082,
  [\href{http://arxiv.org/abs/2204.02990}{{\tt arXiv:2204.02990}}].

\bibitem{Couzens:2022lvg}
C.~Couzens, H.~Kim, N.~Kim, Y.~Lee, and M.~Suh, {\it {D4-branes wrapped on
  four-dimensional orbifolds through consistent truncation}},  {\em JHEP} {\bf
  02} (2023) 025, [\href{http://arxiv.org/abs/2210.15695}{{\tt
  arXiv:2210.15695}}].

\bibitem{Faedo:2022rqx}
F.~Faedo, A.~Fontanarossa, and D.~Martelli, {\it {Branes wrapped on orbifolds
  and their gravitational blocks}},  {\em Lett. Math. Phys.} {\bf 113} (2023),
  no.~3 51, [\href{http://arxiv.org/abs/2210.16128}{{\tt arXiv:2210.16128}}].

\bibitem{Lin:2004nb}
H.~Lin, O.~Lunin, and J.~M. Maldacena, {\it {Bubbling AdS space and 1/2 BPS
  geometries}},  {\em JHEP} {\bf 10} (2004) 025,
  [\href{http://arxiv.org/abs/hep-th/0409174}{{\tt hep-th/0409174}}].

\bibitem{Chong:2004ce}
Z.~W. Chong, H.~Lu, and C.~N. Pope, {\it {BPS geometries and AdS bubbles}},
  {\em Phys. Lett. B} {\bf 614} (2005) 96--103,
  [\href{http://arxiv.org/abs/hep-th/0412221}{{\tt hep-th/0412221}}].

\bibitem{Gutperle:2022pgw}
M.~Gutperle and N.~Klein, {\it {A note on co-dimension 2 defects in N=4,d=7
  gauged supergravity}},  {\em Nucl. Phys. B} {\bf 984} (2022) 115969,
  [\href{http://arxiv.org/abs/2203.13839}{{\tt arXiv:2203.13839}}].

\bibitem{Gutperle:2023yrd}
M.~Gutperle, N.~Klein, and D.~Rathore, {\it {Holographic 6d co-dimension 2
  defect solutions in M-theory}},  \href{http://arxiv.org/abs/2304.12899}{{\tt
  arXiv:2304.12899}}.

\bibitem{Gaiotto:2009gz}
D.~Gaiotto and J.~Maldacena, {\it {The Gravity duals of N=2 superconformal
  field theories}},  {\em JHEP} {\bf 10} (2012) 189,
  [\href{http://arxiv.org/abs/0904.4466}{{\tt arXiv:0904.4466}}].

\bibitem{Romans:1985ps}
L.~J. Romans, {\it {Gauged $N=4$ Supergravities in Five-dimensions and Their
  Magnetovac Backgrounds}},  {\em Nucl. Phys. B} {\bf 267} (1986) 433--447.

\bibitem{Gauntlett:2007sm}
J.~P. Gauntlett and O.~Varela, {\it {D=5 SU(2) x U(1) Gauged Supergravity from
  D=11 Supergravity}},  {\em JHEP} {\bf 02} (2008) 083,
  [\href{http://arxiv.org/abs/0712.3560}{{\tt arXiv:0712.3560}}].

\bibitem{Cassani:2020cod}
D.~Cassani, G.~Josse, M.~Petrini, and D.~Waldram, {\it {$\mathcal{N} $ = 2
  consistent truncations from wrapped M5-branes}},  {\em JHEP} {\bf 02} (2021)
  232, [\href{http://arxiv.org/abs/2011.04775}{{\tt arXiv:2011.04775}}].

\bibitem{Bobev:2022ocx}
N.~Bobev, V.~Dimitrov, and A.~Vekemans, {\it {Wrapped M5-branes and AdS$_{5}$
  black holes}},  {\em JHEP} {\bf 05} (2023) 012,
  [\href{http://arxiv.org/abs/2212.10360}{{\tt arXiv:2212.10360}}].

\bibitem{MatthewCheung:2019ehr}
K.~C. Matthew~Cheung, J.~P. Gauntlett, and C.~Rosen, {\it {Consistent KK
  truncations for M5-branes wrapped on Riemann surfaces}},  {\em Class. Quant.
  Grav.} {\bf 36} (2019), no.~22 225003,
  [\href{http://arxiv.org/abs/1906.08900}{{\tt arXiv:1906.08900}}].

\bibitem{Zaffaroni:2019dhb}
A.~Zaffaroni, {\it {AdS black holes, holography and localization}},  {\em
  Living Rev. Rel.} {\bf 23} (2020), no.~1 2,
  [\href{http://arxiv.org/abs/1902.07176}{{\tt arXiv:1902.07176}}].

\bibitem{Pernici:1984xx}
M.~Pernici, K.~Pilch, and P.~van Nieuwenhuizen, {\it {Gauged Maximally Extended
  Supergravity in Seven-dimensions}},  {\em Phys. Lett.} {\bf 143B} (1984)
  103--107.

\bibitem{Nastase:1999cb}
H.~Nastase, D.~Vaman, and P.~van Nieuwenhuizen, {\it {Consistent nonlinear K K
  reduction of 11-d supergravity on AdS(7) x S(4) and selfduality in odd
  dimensions}},  {\em Phys. Lett.} {\bf B469} (1999) 96--102,
  [\href{http://arxiv.org/abs/hep-th/9905075}{{\tt hep-th/9905075}}].

\bibitem{Nastase:1999kf}
H.~Nastase, D.~Vaman, and P.~van Nieuwenhuizen, {\it {Consistency of the AdS(7)
  x S(4) reduction and the origin of selfduality in odd dimensions}},  {\em
  Nucl. Phys.} {\bf B581} (2000) 179--239,
  [\href{http://arxiv.org/abs/hep-th/9911238}{{\tt hep-th/9911238}}].

\bibitem{Gauntlett:2004zh}
J.~P. Gauntlett, D.~Martelli, J.~Sparks, and D.~Waldram, {\it {Supersymmetric
  AdS(5) solutions of M theory}},  {\em Class. Quant. Grav.} {\bf 21} (2004)
  4335--4366, [\href{http://arxiv.org/abs/hep-th/0402153}{{\tt
  hep-th/0402153}}].

\bibitem{BCtoappear2}
P.~Bomans and C.~Couzens, {\it {Equivariant localisation in the LLM
  background}},  {\em To appear} (2023).

\bibitem{Benini:2012cz}
F.~Benini and N.~Bobev, {\it {Exact two-dimensional superconformal R-symmetry
  and c-extremization}},  {\em Phys. Rev. Lett.} {\bf 110} (2013), no.~6
  061601, [\href{http://arxiv.org/abs/1211.4030}{{\tt arXiv:1211.4030}}].

\bibitem{Benini:2013cda}
F.~Benini and N.~Bobev, {\it {Two-dimensional SCFTs from wrapped branes and
  c-extremization}},  {\em JHEP} {\bf 06} (2013) 005,
  [\href{http://arxiv.org/abs/1302.4451}{{\tt arXiv:1302.4451}}].

\bibitem{Faedo:2019cvr}
A.~F. Faedo, C.~Nunez, and C.~Rosen, {\it {Consistent truncations of
  supergravity and $\frac{1}{2}$-BPS RG flows in $4d$ SCFTs}},  {\em JHEP} {\bf
  03} (2020) 080, [\href{http://arxiv.org/abs/1912.13516}{{\tt
  arXiv:1912.13516}}].

\bibitem{BCtoappear}
P.~Bomans and C.~Couzens, {\it {On the class $\mathcal{S}$ origin of spindle
  solutions}},  {\em To appear} (2023).

\bibitem{Gutperle:2019dqf}
M.~Gutperle and M.~Vicino, {\it {Holographic Surface Defects in $D=5$, $N=4$
  Gauged Supergravity}},  {\em Phys. Rev. D} {\bf 101} (2020), no.~6 066016,
  [\href{http://arxiv.org/abs/1911.02185}{{\tt arXiv:1911.02185}}].

\bibitem{Gutperle:2020rty}
M.~Gutperle and C.~F. Uhlemann, {\it {Surface defects in holographic 5d
  SCFTs}},  {\em JHEP} {\bf 04} (2021) 134,
  [\href{http://arxiv.org/abs/2012.14547}{{\tt arXiv:2012.14547}}].

\bibitem{Gukov:2006jk}
S.~Gukov and E.~Witten, {\it {Gauge Theory, Ramification, And The Geometric
  Langlands Program}},  \href{http://arxiv.org/abs/hep-th/0612073}{{\tt
  hep-th/0612073}}.

\bibitem{BenettiGenolini:2023kxp}
P.~Benetti~Genolini, J.~P. Gauntlett, and J.~Sparks, {\it {Equivariant
  localization in supergravity}},  \href{http://arxiv.org/abs/2306.03868}{{\tt
  arXiv:2306.03868}}.

\bibitem{Martelli:2023oqk}
D.~Martelli and A.~Zaffaroni, {\it {Equivariant localization and holography}},
  \href{http://arxiv.org/abs/2306.03891}{{\tt arXiv:2306.03891}}.

\bibitem{Bah:2018gwc}
I.~Bah and E.~Nardoni, {\it {Structure of Anomalies of 4d SCFTs from M5-branes,
  and Anomaly Inflow}},  \href{http://arxiv.org/abs/1803.00136}{{\tt
  arXiv:1803.00136}}.

\bibitem{Bah:2018jrv}
I.~Bah, F.~Bonetti, R.~Minasian, and E.~Nardoni, {\it {Class $\mathcal{S}$
  Anomalies from M-theory Inflow}},  {\em Phys. Rev.} {\bf D99} (2019), no.~8
  086020, [\href{http://arxiv.org/abs/1812.04016}{{\tt arXiv:1812.04016}}].

\bibitem{Bah:2019jts}
I.~Bah, F.~Bonetti, R.~Minasian, and E.~Nardoni, {\it {Anomaly Inflow for
  M5-branes on Punctured Riemann Surfaces}},
  \href{http://arxiv.org/abs/1904.07250}{{\tt arXiv:1904.07250}}.

\bibitem{Bah:2019rgq}
I.~Bah, F.~Bonetti, R.~Minasian, and E.~Nardoni, {\it {Anomalies of QFTs from
  M-theory and Holography}},  \href{http://arxiv.org/abs/1910.04166}{{\tt
  arXiv:1910.04166}}.

\bibitem{Liu:1999ai}
J.~T. Liu and R.~Minasian, {\it {Black holes and membranes in AdS(7)}},  {\em
  Phys. Lett.} {\bf B457} (1999) 39--46,
  [\href{http://arxiv.org/abs/hep-th/9903269}{{\tt hep-th/9903269}}].

\bibitem{Gauntlett:2006ai}
J.~P. Gauntlett, E.~O~Colgain, and O.~Varela, {\it {Properties of some
  conformal field theories with M-theory duals}},  {\em JHEP} {\bf 02} (2007)
  049, [\href{http://arxiv.org/abs/hep-th/0611219}{{\tt hep-th/0611219}}].

\end{thebibliography}\endgroup

\end{document}